\documentclass[aps,prb,singlecolumn,groupedaddress,showpacs,superscriptaddress,amssymb,amsmath]{revtex4-1}
\usepackage[version=3]{mhchem} 
\usepackage[T1]{fontenc}       

\usepackage{graphicx}
\usepackage{tabularx}
\usepackage{color}
\usepackage{amsmath}
\usepackage{comment}
\usepackage{ifthen}
\newcommand{\be}{\begin{equation}}
\newcommand{\ee}{\end{equation}}
\newcommand{\bea}{\begin{eqnarray}}
\newcommand{\eea}{\end{eqnarray}}
\usepackage{dcolumn}
\usepackage{hyperref}
\usepackage{bm}
\usepackage{epstopdf}
\usepackage{amsfonts}
\usepackage{tikz}

\newcommand\tikzmark[1]{%
        \tikz[overlay,remember picture,baseline] \node [anchor=base] (#1) {};}

\newcommand\MyLine[3][]{%
        \begin{tikzpicture}[overlay,remember picture]
        \draw[#1] (#2.north west) -- (#3.south east);
\end{tikzpicture}}

\newcommand{\kr}[1]
        {
        \ifthenelse{\equal{#1}{no}}
        {\overrightarrow{k}}
        {\overrightarrow{k_{#1}}}
        }
\newcommand{\kl}[1]
        {
        \ifthenelse{\equal{#1}{no}}
        {\overleftarrow{k}}
        {\overleftarrow{k_{#1}}}
        }

\begin{document} 

\title{On the relationship between the mean first-passage time and the steady state transfer rate in classical chains}  
\author{Na'im Kalantar}
\affiliation{
Department of Chemistry and Centre for Quantum Information and Quantum Control,
University of Toronto, 80 Saint George St., Toronto, Ontario, Canada M5S 3H6
}

\author{Dvira Segal}
\affiliation{
Department of Chemistry and Centre for Quantum Information and Quantum Control,
University of Toronto, 80 Saint George St., Toronto, Ontario, Canada M5S 3H6
}
\email{dvira.segal@utoronto.ca}




\begin{abstract}
Understanding excitation and charge transfer in disordered media is a significant
challenge in chemistry, biophysics and material science.
We study two experimentally-relevant measures  
for carriers transfer in finite-size chains, the trapping
mean first-passage time (MFPT) and the steady state transfer time (SSTT).
We discuss the relationship between these measures, 
and derive analytic formulae for one-dimensional chains.
We exemplify the behavior of these timescales in different motifs: 
donor-bridge-acceptor systems, biased chains,
and alternating and stacked co-polymers.
We find that the MFPT and the SSTT may administer different, 
complementary information on the system, jointly reporting on molecular length and energetics.
Under constraints such as fixed donor-acceptor energy bias, 
we show that the MFPT and the SSTT are optimized (minimized)
under fundamentally different internal potential profiles.
This study brings insights on the behavior of the MFPT and the SSTT, and suggests that it is beneficial
to perform both transient and steady state measurements on a conducing network
so as to gather a more complete picture of its properties.
\end{abstract}

\maketitle

\section{Introduction}

Understanding charge and energy transfer processes in soft and disordered materials such as thin films made from
organic polymers or large biological molecules is central to the development of 
electronic and energy solar-harvesting applications \cite{VMay}.
To model such systems efficiently, a coarse-grained model of the surrounding environment is often assumed.
In turn, this approach allows the application of kinetic, Markovian master equations for 
following the dynamics of averaged variables. 

Mean first-passage quantities are useful for characterizing stochastic processes.
Specifically, the mean first-passage time defines a timescale 
to visit a specified target (or a threshold value) for the first time \cite{VKampen}. 
This measure finds numerous applications in physics, chemistry and biology, 
e.g. for quantifying reaction rates,
transmission of particles in channels, molecular processes such as receptor binding and adhesion,
and cellular processes such as cell division \cite{Zilman}.
Here, we focus on (charge or energy) transfer processes in flexible molecular systems 
and use the mean first-passage time to report on the transfer process from a certain initial site to a final location \cite{Klafter98}.

The capacity of molecules to transfer electric charge can be measured in different setups.
In transient absorption spectroscopy experiments, an initial state is carefully prepared and monitored in time,
see e.g. Refs. \cite{LewisJacs,LewisAcc,Majima}.
In contrast, in electrochemical experiments a molecule of interest is attached to an electrode,
and the rate of charge transfer from the electrode to redox groups in the molecule is monitored, 
see e.g. Ref. \cite{Waldeck}.
Alternatively, in molecular conductance experiments a molecule links
two voltage-biased electrodes,  e.g., an STM tip and a conducting substrate, 
and the molecule is characterized by its electrical conductance \cite{BJ}.
To thoroughly understand charge transfer processes in a single molecule or ensemble of molecules, 
it is critical to resolve the connection between measured observables 
in such different experiments.

Several theoretical studies predicted a linear relationship between the intramolecular charge transfer rate 
in a donor-bridge-acceptor system and the low bias conductance in a metal-molecule-metal junction 
\cite{NitzanR1,NitzanR2,Berlin,Traub,BeratanP}.
Nevertheless, experiments revealed a more complex behavior \cite{Zhou,Waldeck,Bueno}.
For example, in Ref. \cite{Waldeck} the electrochemical rate constant
and the molecular conductance were examined in alkane chains and peptide nucleic acid oligomers, 
generally manifesting nonlinear correlations between the transfer rate and the conductance.
These deviations from linearity are rationalized by noting that, for the same molecule, transient, 
electrochemical, and conductance measurements 
are performed at undoubtedly different settings, considering ensemble vs. single molecule,  
using different solvents, and experiencing different environmental conditions.
As a result, the energetics of the molecule and its dephasing and relaxation processes are different in the 
these distinct types of experiments.

In this work, we revisit the following basic question: 
Considering carrier transfer processes in one-dimensional hopping systems,
what is the relationship between transient measures and steady state observables,
particularly going beyond donor-bridge-acceptor systems?
We leave aside the challenging, practical problem discussed above, that in different experimental
settings the molecule and its local environment are modified, and for simplicity assume that the 
structures are identical in the different types of experiments.

Using classical rate equations, we compare transient and steady state measures for the transfer process,
namely, the trapping mean first-passage time (MFPT) and the inverse of the steady state rate, 
referred to as the steady state transfer time (SSTT).
Based on these two measures, we study the roles of structural motifs 
on carrier transfer in different networks, including
linear or branched chains, uniform, bridge-mediated, energy-biased, single and multi-component networks.
We derive a simple, intuitive relationship between the two timescales, MFPT and the SSTT, and
explain under what conditions they agree.
Generally, we find that these measures scale differently with system size and energy,
and that they can disclose distinct properties of the system. 
We further discuss the enhancement of the transfer speed by optimizing the 
energy profile of the system, achieved e.g. by chemical modifications or gating.

\section{Model and Measures} 
\label{Model}

We study the problem of a random walk across a finite system. 
The model includes $n+2$ sites, with a donor D (site 0), acceptor A (site $n+1$), and $n$ intermediate sites.
In addition, the model includes a trap $T$;  transitions from A to the trap are irreversible 
with a trapping rate constant $\Gamma_A$.
Variants of this model have been adopted to describe charge transfer and excitation energy transport in 
polymers, biomolecules, and amorphous systems.  A model with multiple acceptors is discussed in the Appendix. 
At this point, we do not specify the connectivity nor the energy profile of the system, 
to be described in examples. As well, we do not include lossy processes within the system, 
such as exciton recombination. 

The Pauli master equation describes the time evolution of sites populations, $p_{i}$, $i=0,1,..,n+1$,
\bea
\dot {\bf p}(t) = {\bf M p}(t),
\label{eq:psol}
\eea
with the vector of population {\bf p} and rate constant matrix ${\bf M}$. 
Note that this system of equations does not include the equation of motion for the trap, which
fulfills $\dot p_T= \Gamma_A p_A$.
In such a coarse-grained random walk model, the properties of the environment, 
such as its temperature, determine the hopping rates; 
to make a transition up (down) in energy the walker absorbs (dissipates) heat from (to) the 
surrounding thermal bath.
The coupled kinetic rate equations (\ref{eq:psol}) are linear and first-order, and therefore can be readily solved.  
We particularly mention that in many instances, 
experiments of charge transfer in DNA are well described by the Pauli rate equations, see e.g.  
\cite{Bixon,Giese,Berlin01,Burin,Porath,Wagner}.

The success rate of a transfer process in a kinetic network can be quantified 
using different measures, such as the transfer time, yield, flux.
Before we introduce the different observables, 
we present time dependent and steady state experiments.
In the first case, realized e.g. with transient absorption spectroscopy,
one prepares a well-defined initial condition and follows the time evolution of the system towards 
its long-time state. For example, in studies of charge transfer in DNA, an excess charge is prepared on a 
donor site using a laser excitation. Once injected into the DNA, the excess charge migrates until 
it arrives at an acceptor molecule attached to the DNA  \cite{LewisJacs,LewisAcc,Majima}. 
Mathematically, transient dynamics is revealed by solving the time dependent, 
first order differential equation
(\ref{eq:psol}). The formal solution is ${\bf p}(t) =\exp ({\bf M}t) {\bf p_0}$, with the vector of the initial condition 
${\bf p_0}$. 

Alternatively, one may study the same system---albeit in a steady state situation---by defining
a boundary condition rather than an initial condition, identifying a source and a sink 
(possible more than one). 
For example, the donor population can be maintained fixed by continuously feeding in particles; 
population leaves the acceptor at the same rate.
In this scenario, the relevant measure for the transfer process is the flux of outgoing particles, 
from the acceptor---outside.
As was discussed in e.g. Refs. \cite{NitzanR1,NitzanR2,Berlin,Traub,BeratanP}, in a donor-bridge-acceptor 
configuration and under some simplifying assumptions this steady state setup 
can be furthermore related to the low-voltage molecular conductances.

In what follows, we consider time dependent and steady state settings, 
examine transfer measures, and discuss their relationships.
We emphasize that we focus here on systems that include a single donor and a single acceptor,
and that so far we do not consider losses within the network.

\subsection{Mean first-passage time}
\label{secMFPT}

The mean first-passage time defines a timescale for a random event to occur for the first time.
A trapping timescale  describes the mean time it takes for
population to leave the system, that is, to transfer from the donor to the trap.
Following Ref. \cite{BeratanIs}, we define the trapping mean first-passage time as
\bea
\tau_m \equiv \int_0^{\infty} t \dot p_T(t) dt.
\label{eq:MFPTdef}
\eea
Here, $p_T(t)$ is the trap population and 
$\dot p_T(t)$ is the probability density of the passage time. 
Since the probability of not having transitioned to the
trap is equal to the probability of being in the other $n+2$ sites of the system, 
$\dot p_{T}(t) = -\sum_{j=0}^{n+1} \dot p_j(t)$. 
Integrating by parts we get
\bea
\tau_m &=&\sum_{j=0}^{n+1} \int_0 ^{\infty}p_j(t) dt 
\nonumber\\
&=& \sum_{j=0}^{n+1}r_j,
\label{eq:MFPT1}
\eea
where the residence time is defined as
\bea
r_j\equiv \left[\int_0^{\infty} \exp({\bf M} t) {\bf p_0} dt\right]_j = \left[-{\bf M}^{-1} {\bf p_0}\right]_j.
\label{eq:residencet}
\eea
Note that we count rows starting from 0 to $n+1$. 
Alternatively, the MFPT can be calculated from the accumulated population on the acceptor; the rate equation describing the kinetics of the trap fulfills
$\dot p_T(t) = \Gamma_Ap_A(t)$, therefore
\bea
\tau_m = \Gamma_A \int_0^{\infty} t p_A(t) dt.
\label{eq:MFPT2}
\eea
Eqs.\@ (\ref{eq:MFPT1}) and (\ref{eq:MFPT2}) are equivalent; computationally, the former is more convenient. We emphasize that the MFPT, as well as other
observables considered below, are calculated without the explicit consideration of the trap state population.

Another method to calculate the MFPT is to replace the trap and the donor states 
with a combined donor/trap state. The steady state for this configuration gives the MFPT by 
\bea
\frac{1}{\tau_m} = \Gamma_A  p_{D/T},
\eea
where $p_{D/T}$ is the steady state population of the combined donor/trap state. 
This scheme allows calculation of MFPT for systems with a backflow from the trap state to the acceptor \cite{BeratanFDisc}.
In the rest of the paper, we calculate the MFPT based on Eq. (\ref{eq:MFPT1}).


\subsection{Mean transfer time}

Another measure that has been commonly introduced in the literature is the average transfer time to 
reach the acceptor, written as
\bea
\tau_t\equiv \frac{\int_0^{\infty} t p_A(t) dt }{\int_0^{\infty}p_A(t)dt }.
\label{eq:Mt}
\eea
Nevertheless, since   $\int_0^\infty \Gamma_A p_A dt =  \int_0^\infty  \dot p_T dt=1$, 
Eq.\@ (\ref{eq:Mt})  is identical to Eq.\@ (\ref{eq:MFPT2}), concluding that the MFPT and the mean transfer time are in fact the same observable here. 
Note that these considerations hold even if the underlying dynamics involves nonthermal effects and/or quantum effects
comprising the time evolution of coherences.

\subsection{Steady state flux}
\label{secSSTT}

In a steady state experiment we maintain the population of the donor state fixed, $p_D(t)=1$, and 
measure the outgoing particle flux from the acceptor towards the trap.
In the long time limit, transient effects die out, $\dot {\bf p}(t)=0$, 
and the problem reduces to an algebraic equation,
\bea
 \widetilde {\bf M} {\bf p}_{ss}= {\bf v}.
\eea
Here, $\widetilde {\bf M}$ corresponds to the matrix ${\bf M}$ of Eq. (\ref{eq:psol}) by replacing its first row by
the condition $p_D^{ss}=1$, which is the donor state population in steady state. The inhomogeneous column ${\bf v}$ has a single nonzero value (unity) in its first element. 
The steady state flux is defined as
\bea
k_{ss}&\equiv&\Gamma_A p_A^{ss}/p_D^{ss}
\nonumber\\
&=& \Gamma_A\left[\widetilde {\bf M}^{-1}{\bf v} \right]_{n+1}. 
\label{eq:kss}
\eea
The inverse of the flux defined a timescale
\bea
\tau_{ss}\equiv k_{ss}^{-1},
\label{eq:SSTT}
\eea
which is referred to as the ``steady state transfer time''.
%

\subsection{On the relation between the MFPT and the steady state flux}
\label{Sec-relation}

Since different protocols (transient and steady state) may be used to interrogate the same system,
it is critical to resolve the relationship between the MFPT and the SSTT. 
Do these measures necessarily convey the same information? 
If not, when do these timescales agree? 
This question will be discussed throughout the paper: We provide examples in which the 
two measures, (\ref{eq:MFPT1}) and (\ref{eq:SSTT}) converge, and situations in which they report on
different properties of the system.

Intuitively, as was explained in e.g. Ref. \cite{nitzancar},
the MFPT and the SSTT agree when the dynamics is dominated by a single decaying exponent. 
Considering the eigenvalues of the kinetic matrix ${\bf M}$, this situation manifests itself when
$|(\lambda_1-\lambda_0)/\lambda_0|\gg1$. Here,
 $\lambda_0$  is the smallest (in magnitude) eigenvalue of ${\bf M}$, and $\lambda_1$ is the next eigenvalue in order 
of increasing magnitude. 
While mathematically, this condition is clear, its physical implications are not transparent. In what follows, 
we derive a relationship between the MFPT and the SSTT, which can be readily rationalized.

First, we note that in the time dependent protocol, as explained in Sec. \ref{secMFPT}, 
the total population of the system  ($n+2$ sites and trap) adds up to one. 
In contrast, in a steady state experiment as described in Sec. \ref{secSSTT}, 
the population of the system  exceeds unity.
Let us now define the rate constant 
$k_{ss} = \Gamma_A p_A^{ss}$ [we ignore thus far the denominator of Eq.\@ (\ref{eq:kss})].
Since the outgoing flux is equal to the incoming flux, we can write down
%
\bea {\bf M} {\bf p}^{ss} = \begin{pmatrix} - k_{ss} & 0 & 0 & \dots \end{pmatrix}^T = - k_{ss}{\bf p}_0,
\label{eq:popss}
\eea
with ${\bf p}^{ss}$  the vector of population in steady state,
which can be obtained by inverting ${\bf M}$. 
Using Eq.\@ (\ref{eq:residencet}), the solution to Eq. (\ref{eq:popss}) is given in terms of the residence times,
\bea
{\bf  p}^{ss} =k_{ss} {\bf r}.
\eea
Summing up the rows and identifying the MFPT by the sum of the residence times we get
\bea
\sum_{j=0}^{n+1} [ p^{ss}]_j =   k_{ss} \tau_m.
\eea
We now define the total steady state population of the system as $p^{ss}_{sys}\equiv \sum_{j=0}^{n+1} [ p^{ss}]_j$,
and arrive at
\bea
\tau_{ss} p_{sys}^{ss}= \tau_m.
\eea
This simple  expression directly relates the MFPT to the SSTT, and it hands over a physical meaning to the eigenvalue argument 
mentioned above. It is important to note that this relation is general and that it does not rely on the details of the network
i.e., the model may include transitions beyond nearest neighbors, loops, etc. 

We now consider different normalizations for the steady state populations:
(i) A condition which is not straightforward to experimentally implement is
$\sum_{j=0}^{n+1} p_{j}^{ss}=1$,  that is, the total population of the system is maintained at unity
in the steady state limit. In this (unnatural) case, the SSTT and the MFPT precisely coincide,
$\tau_{ss} = \tau_m$.
(ii) Alternatively,  we can set the steady state 
by invoking the more plausible choice, $p_D^{ss}=1$, which is the common case discussed 
in Sec. \ref{secSSTT}.  We then find that
the total population, $\sum_{j=0}^{n+1} p_{j}^{ss}$, exceeds unity and therefore the 
two timescales diverge. 
Nevertheless, as long as $\sum p_{j\neq D}^{ss} \ll 1$, the total 
population of the system remains close to 1,  and therefore the SSTT approaches the MFPT.
This scenario precisely arises in a donor-bridge-acceptor model when the energy of the bridge is 
placed high relative to the energy of the donor  \cite{Segal00}. 


\section{One-dimensional chains with arbitrary rates: Analytic results and examples}
\label{sec:1D}

In this section, we derive analytical expressions for the MFPT and the SSTT in a 
nearest-neighbor one-dimensional (1D) model.
We then exemplify these measures in problems that are 
interesting in the context of charge and excitation transfer in
molecular chains and junctions such as organic polymers and DNA.

The model includes  a single donor, $n$ intermediate sites on a chain, an acceptor, and a trap.  We calculate the MFPT and the SSTT, 
discuss their relationship, and exemplify that they may convey distinct, complementary information.
Our analytical expression for the MFPT agrees with a previous work \cite{Klafter98}. Nevertheless,
our simple derivation and the associated analysis of the SSTT
provide new physical insights.
The population equations of motion satisfy  ($i=1,2...,n$)
\bea
&&\dot p_0(t) =-\kr{1}p_0(t) + \kl{0} p_1(t),
\nonumber\\
&&\dot p_i(t) =-\left( \kl{i+1}+\kr{i-1} \right) p_i(t) + \kl{i} p_{i+1}(t) +\kr{i}p_{i-1}(t), 
\nonumber\\
&&\dot p_{n+1}(t) = -\left( \kl{n} + \Gamma_A \right)  p_{n+1}(t) 
+ \kr{n+1} p_{n}(t),
\label{eq:kin}
\eea
with $\kl{i}$ ($\kr{i}$) as the rate constant to make a transition towards site $i$ 
which is at the left (right) of the current site.
For example, assuming an equilibrium bath at temperature $T$, 
the rate constants satisfy the detailed balance relation, 
$\frac{\kr{i+1}}{\kl{i}}=e^{-(E_{i+1}-E_i)/T}$; 
we set the Boltzmann constant as $k_B\equiv 1$.
Nevertheless, our general results in this Section do not rely on this particular choice for the rates.


\subsection{MFPT}
\label{3MFPT}

The rate matrix (\ref{eq:kin}) of dimension $(n+2)\times(n+2)$ is tridiagonal (all missing entries are zero),
%
\bea \label{eq:arbratematrix}
\textbf{M} = \begin{pmatrix}
-\kr{1}          & \kl{0}          &\tikzmark{c}      &             &                   &  \\ 
\kr{1}           &-\kl{0}-\kr{2}   & \kl{1}           &             & \mbox{\Huge 0}                  &                   \\ 
\tikzmark{a}     & \kr{2}          &-\kl{1}-\kr{3}    &\ddots       &                   &                   \\ 
                 &                 &\kr{3}            &\ddots       & \kl{n-1}          &\tikzmark{d}       \\ 
                 &\mbox{\Huge 0}&                &\ddots       & -\kl{n-1}-\kr{n+1}& \kl{n}            \\ 
 &                &                  & & \tikzmark{b} \kr{n+1}         &-\kl{n} - \Gamma_A \\ 
\end{pmatrix} 
\eea
%
Our objective is to find the residence times using
Eq.\@~(\ref{eq:residencet}), $-\textbf{M}\left({\bf r}\right) = {\bf p}_0$.
Since  ${\bf M}$ is tridiagonal, by following the Thomas algorithm \cite{tridiagonal}
we reduce the above equation to   
%
 \bea 
 \begin{pmatrix}
 1 & -\frac{\kl{0}}{\kr{1}} &\tikzmark{g}            &       & \mbox{\Huge 0} \\
\tikzmark{e}   & 1          & -\frac{\kl{1}}{\kr{2}} &       &  \\
   &                        & 1                      &\ddots &\tikzmark{h}  \\
   &                        &                        &\ddots &-\frac{\kl{n}}{\kr{n+1}}  \\
\mbox{\Huge 0}&                        &                        &\tikzmark{f}       &  1   \\
\end{pmatrix} 
\left( {\bf r} \right) = 
  \begin{pmatrix} 
\kr{1}^{-1} \\\kr{2}^{-1} \\\kr{3}^{-1} \\ \vdots \\ \kr{n+1}^{-1} \\ \Gamma_A^{-1}
\end{pmatrix} 
 \eea
\MyLine[thick]{e}{f}
\MyLine[thick]{g}{h}
The transfer times $r_j$ are obtained by progressing recursively from the last row to the first row,
\bea
{\bf r} = 
\begin{pmatrix}
\frac{1}{\kr{1}} \left( 1 + \kl{0}  r_1 \right) \\  
\frac{1}{\kr{2}} \left( 1 + \kl{1}  r_2 \right) \\  
\frac{1}{\kr{3}} \left( 1 + \kl{2}  r_3 \right) \\  
\vdots \\
\frac{1}{\kr{n+1}} \left( 1 + \kl{n}  r_{n+1} \right)  \\
\frac{1}{\Gamma_A}
\end{pmatrix}
\label{eq:arbres}
 \eea
or explicitly,
\bea
{\bf r} =
 \begin{pmatrix}
\frac{1}{\kr{1}} \left(1 + \frac{\kl{0}}{\kr{2}}\left( 1 + \frac{\kl{1}}{\kr{3}} + \dots \right) \right)\\
\frac{1}{\kr{2}} \left(1 + \frac{\kl{1}}{\kr{3}}\left( 1 + \frac{\kl{2}}{\kr{4}} + \dots \right) \right)\\
\frac{1}{\kr{3}} \left(1 + \frac{\kl{2}}{\kr{4}}\left( 1 + \frac{\kl{3}}{\kr{5}} + \dots \right) \right)\\
\vdots \\
\frac{1}{\kr{n+1}} + \frac{\kl{n}}{\kr{n+1}\Gamma_A} \\
\frac{1}{\Gamma_A}
\end{pmatrix}
\eea
The recursive form has a clear interpretation: $\kr{i+1}^{-1}$  
is the residence time on site $i$ before moving forward once;
$r_{i+1}\kl{i}/ \kr{i+1} $ is the contribution to the $i$th residence time from a particle
that jumps to site $i$ from $i+1$, before moving forward again.
The mean first-passage time is the sum of the residence times, given by
\bea
\tau_m = 
\frac{1}{\kr{1}} \left(1 + \frac{\kl{0} + \kr{1}}{\kr{2}}\left( 1 + \frac{\kl{1} + \kr{2}}{\kr{3}} \left( \dots \right) \right) \right)
\label{eq:arbmftt}
\eea
The  recurrence relation for the residence time can also be seen from the flux condition: 
The population must leave to the right 
one more time than it enters from the left,
$ \kr{i} r_{i-1} - \kl{i-1} r_{i} = 1$.
  

\subsection{SSTT}
\label{3SSTT}

Following the discussion of Sec. \ref{secSSTT}, our objective is to solve 
the algebraic equation $\widetilde {\bf M}{\bf p}^{ss} = {\bf p}_0$. We manage to reduce it as follows,
%
 \bea  
\label{ArbKss}
 \begin{pmatrix} 
 1          & 0 &\tikzmark{k}         &                           &       &             & \\
\tikzmark{i}& 1 &-\frac{\kl{1}}{q(1)} &                           & \mbox{\Huge  0}& & \\
            &   & 1                   & -\frac{\kl{2} q(1)}{q(2)} &       &             &  \\
            &   &                     & 1                         & \ddots&             & \tikzmark{l}\\
            &   &\mbox{\Huge  0}   &                           & \ddots&1            & -\frac{\kl{n-1}q(n-2)}{q(n-1)} \\
            &   &                     &                           &       &\tikzmark{j} & 1 \\ 
 \end{pmatrix}
 {\bf p}^{ss} = 
 \begin{pmatrix} 
 1 \\
 \left( \frac{1}{\kr{1}} + \frac{\kl{0}}{\kr{1}\kr{2}} \right)^{-1} \kr{2}^{-1} \\
 \left( \frac{1}{\kr{1}} + \frac{\kl{0}}{\kr{1}\kr{2}} + \frac{\kl{0}\kl{1}}{\kr{1}\kr{2}\kr{3}} \right)^{-1} \kr{3}^{-1}  \\
 \vdots \\
 \left( \frac{1}{\kr{1}} + \dots + \frac{\prod_{i=0}^{n-1}\kl{i}}{\prod_{i=1}^{n+1}\kr{i}} \right)^{-1} \kr{n+1}^{-1}  \\
 \left( \frac{1}{\kr{1}} + \dots + \frac{\prod_{i=0}^{n}\kl{i}}{\Gamma_A \prod_{i=1}^{n+1}\kr{i}} \right)^{-1} \Gamma_A^{-1} 
 \label{eq:ssTTc}
 \end{pmatrix}
\eea	
\MyLine[thick]{i}{j}
\MyLine[thick]{k}{l}
Here, $q(1) = \kr{2} + \kl{0}$, $q(2) = \kr{2}\kr{3} + \kr{3} \kl{0} + \kl{0} \kl{1}$, 
$q(3)=\kl{2}\kl{0}\kl{1} + \kr{4} \kr{2}\kr{3} + \kr{4} \kr{3}\kl{0}+\kr{4}\kl{0}\kl{1}$, and so on.
To generate these combinations, 
we define the sequence $s(x)$=$\left\{\kr{2}, \dots, \kr{x+1}, \kl{0}, \kl{1}, \dots, \kl{x-1} \right\}$;
 $q(x)$ is a sum of $x+1$ products of nearest neighbors elements in the sequence $s(x)$.
For convenience, we use the notation  $\kr{n+2} = \Gamma_A$. 
	
The steady state flux is given by $\Gamma_A p_{n+1}$, which can 
be found directly from the last row of Eq.\@ (\ref{ArbKss}). 
The other elements of the column vector on 
the right-hand side of Eq.\@ (\ref{ArbKss}) are the steady state populations 
for chains that end with the site corresponding to that row.

\begin{figure}[htbp]
\includegraphics[scale=0.3]{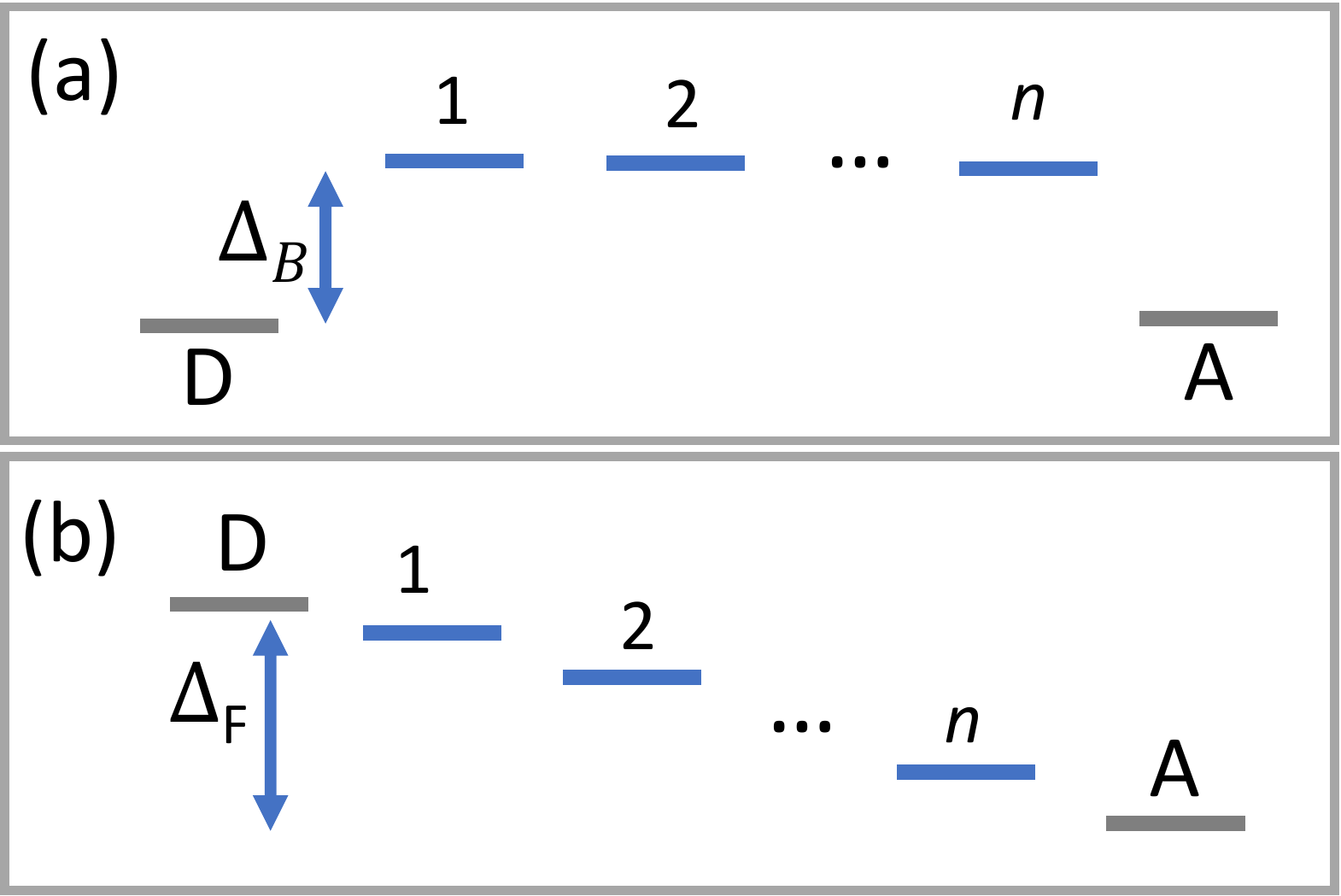} 
\caption{One-dimensional models:
(a) Donor-bridge-acceptor chain with an energy gap (scaled by the temperature) $\Delta_B$.
(b) Biased system, exemplifying here a potential bias $\Delta_F$ towards the acceptor.
}
\label{Figs1}
\end{figure}

\begin{figure}[htbp]
\includegraphics[scale=0.28]{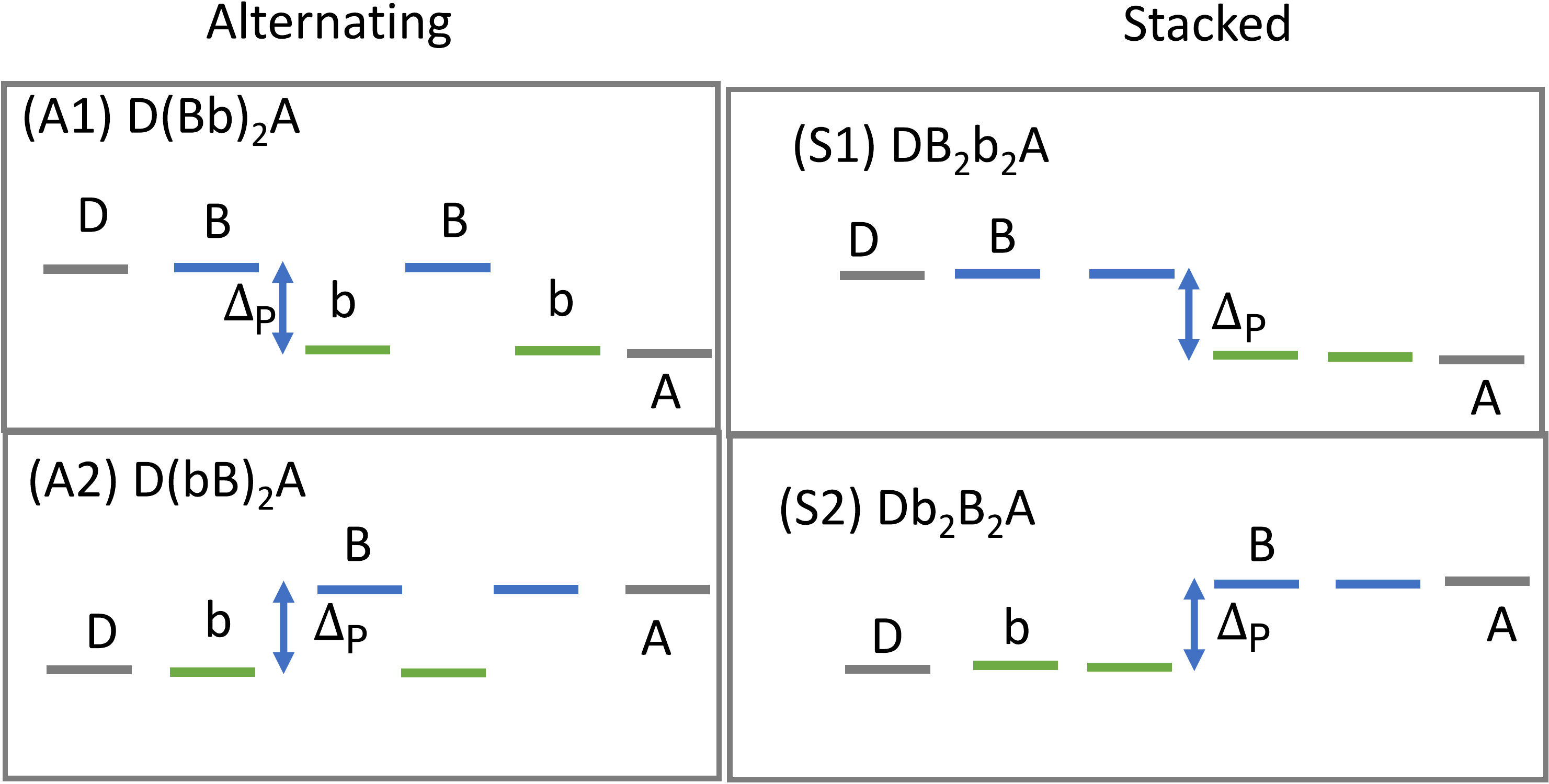} 
\caption{Examples of co-polymer motifs.
(A1)-(A2) Alternating chains with D connected either to the high or low energy monomers.
(S1)-(S2) Stacked chains with D connected either to the high or low energy monomers.
The D (A) sites are placed in resonance with the first (last) sites.
There are even number of sites in the system in the presented examples. In the odd case,
an additional site is added right before the acceptor site. 
}
\label{Figs2}
\end{figure}

In what follows, we exemplify our results for the MFPT and SSTT on several models, 
see Figs.\@ \ref{Figs1}-\ref{Figs2}:
(i) Donor-bridge-acceptor configuration, 
where the bridge's energy are set constant and higher than the D and A levels.
(ii) Biased chain, representing a system under a constant electric field.
(iii) Co-polymer motifs, either alternating or stacked, 
representing charge transport in quasi 1D chains such
as motion along the pi stacking in a double stranded DNA.
Our results in these three cases assume that transitions are induced by a heat bath at 
thermal equilibrium,
therefore we enforce the detailed balance relation. We work with
dimensionless energy parameters, scaled by temperature.

\subsection{Example I: Donor-Bridge-Acceptor system}

We consider here the ubiquitous donor-bridge-acceptor setup. 
We assume that bridge energies are uniform, and that they are 
higher than both the donor and the acceptor levels, see Fig. \ref{Figs1}(a).
The MFPT is calculated from Eq.\@  (\ref{eq:arbres}). Enforcing detailed balance we set 
 $\kr{1} = e^{-\Delta_B}$, $\kl{n} = e^{-\Delta_B}$, and all other $\kr{i}, \kl{i} = 1$;
 $n$ is the number of bridge sites and $\Delta_B$ is the dimensionless energy gap between the donor/acceptor and the bridge, (scaled by the temperature). 
The residence times are given by
\bea \bf{r} = 
\begin{pmatrix} 
		\frac{1}{\Gamma_A} + (n+1)e^{\Delta_B} \\  
		\frac{e^{-\Delta_B}}{\Gamma_A} + n \\ 
		\frac{e^{-\Delta_B}}{\Gamma_A} + (n - 1) \\ 
		\frac{e^{-\Delta_B}}{\Gamma_A} + (n - 2) \\ \vdots \\ 
		\frac{e^{-\Delta_B}}{\Gamma_A} + 1 \\ 
		\frac{1}{\Gamma_A} 
\end{pmatrix}
\eea 
Summing them up while setting for simplicity $\Gamma_A = 1$ we get
\bea
\tau_m = ne^{-\Delta_B} + (n+1)e^{\Delta_B} + \frac 12 n^2 + \frac 12 n + 2.
\label{eq:bridgeMFTT} 
\eea
When $\Delta_B =0$, this formula reduces to  known results
$\tau_m =(n+2)(n+3)/2 $,  see  \cite{BeratanIs}.
In the opposite limit, when $\Delta_B$ is large, the linear term dominates Eq.\@ (\ref{eq:bridgeMFTT}),
and it approaches the scaling $\tau_m \propto (n+1)e^{\Delta_B}$.
 
The steady state transfer time is derived from the last row of Eq.\@ (\ref{ArbKss}),
\bea 
k^{-1}_{ss} &=& 
\frac{1}{\kr{1}} + \dots + \frac{\prod_{i=0}^{n}\kl{i}}{\prod_{i=1}^{n+1}\kr{i}} 
\nonumber\\
&=& 
e^{\Delta_B} + \dots + e^{\Delta_B} + 1, 
\eea
resulting in
\bea 
\tau_{ss} = (n+1)e^{\Delta_B} + 1.  
\label{KSS} 
\eea 
This measure is linear in $n$ for all $\Delta_B$. 
We therefore conclude that the MFPT and the SSTT agree so long as 
$\Delta_B \gg 1$. In this case, the bridge population is small,  
which agrees with our analysis of Sec. \ref{Sec-relation}.

\begin{figure}[h]
\includegraphics[scale=0.34]{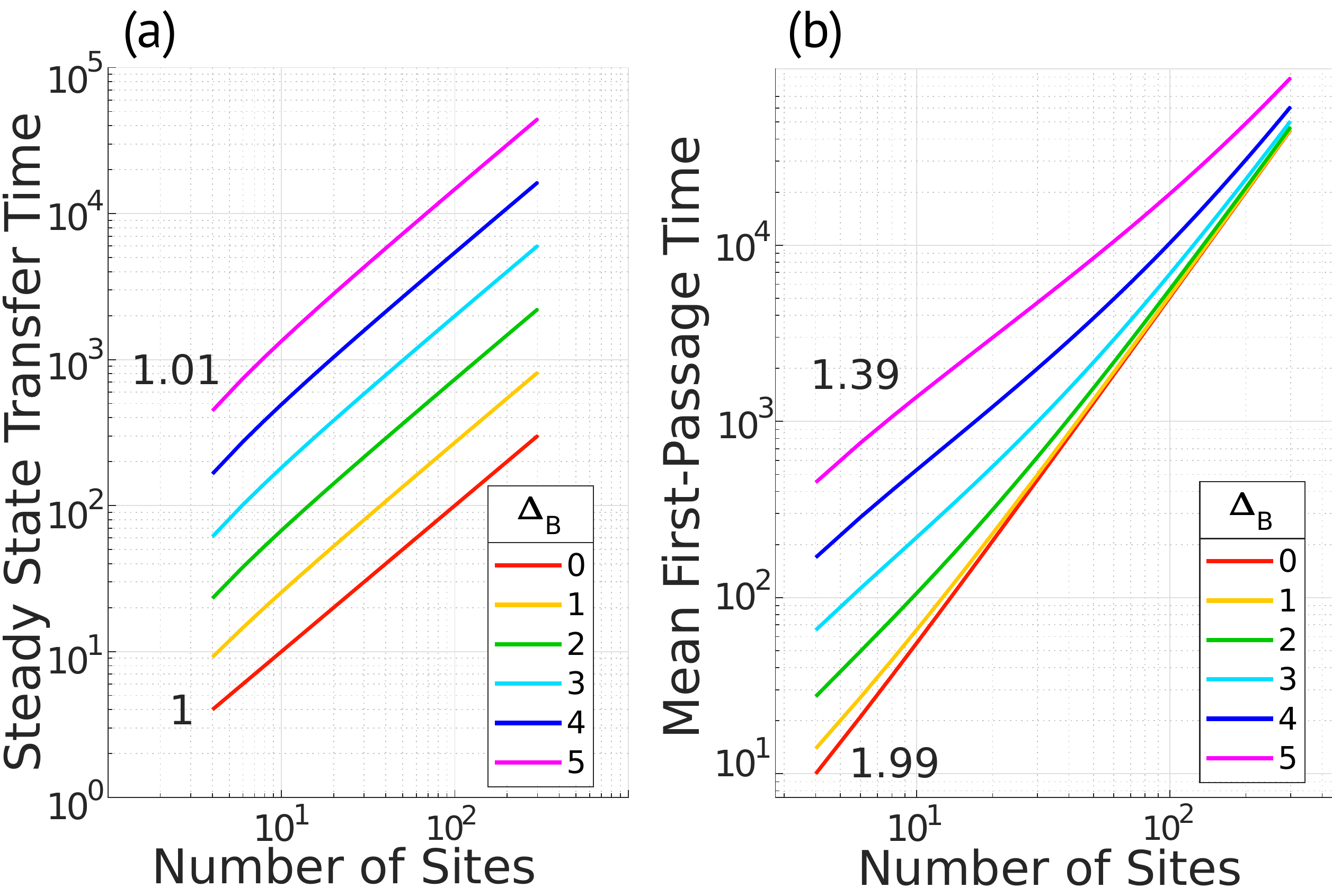} 
\caption{Dependence of (a) steady state transfer time 
and (b) mean first-passage time on $n$ for  donor-bridge-acceptor systems.}
\label{fig-B}
\end{figure}

In Fig.\@ \ref{fig-B}, we display the MFPT and the SSTT while 
increasing the system size and the bridge height $\Delta_B$.
The SSTT scales linearly with $n$,
while the MFPT shows a transition from  quadratic to linear scalings upon the  increase in $\Delta_B$.

\subsection{Example II: Biased chain} 
We consider now a biased chain, as illustrated in Fig. \ref{Figs1}(b).
The system can be biased in opposite polarities, such that energies decrease or increase towards the acceptor.
In what follows, we consider both cases, which we refer to as negative (decreasing) and positive (increasing) potential bias.

{\it Negative potential bias, $E_D>E_A$.}
The donor energy is placed above the acceptor, and the total gap is denoted by $\Delta_F\equiv |E_D-E_A|$.
Recall that the chain comprises the donor, acceptor, and $n$ intermediate states. 
For a non-increasing potential bias,
we get from Eq. (\ref{eq:arbmftt}) the MFPT
\bea
\tau_m &=& (n+2) + \sum_{i=0}^n \kl{i}
+ \sum_{i=1}^n \left( \kl{i} \kl{i-1}\right)
\nonumber\\
&+& \sum_{i=2}^n \left( \kl{i} \kl{i-1} \kl{i-2}\right) + \dots + \left(\kl{0}\kl{1}\dots\kl{n} \right). 
\label{eq:descendmftt}
\eea
We now specifically assume a linear, constant gradient profile such that
 $\kl{no} = e^{-\frac{\Delta_F}{n+1}}$  and $\kr{no} = 1$.
Using the analytical results of Secs. \ref{3MFPT} and \ref{3SSTT}, we find  the MFPT
\bea 
\tau_m &=& (n+2) + (n+1)e^{-\frac{\Delta_F}{n+1}}
	+ ne^{-2\frac{\Delta_F}{n+1}}
\nonumber\\
	&+& (n-1)e^{-3\frac{\Delta_F}{n+1}} 
	+ \dots + e^{-\Delta_F},
\eea
and the SSTT  
\bea 
	\tau_{ss} = 1+ 
	 e^{-\frac{\Delta_F}{n+1}}
	+ e^{-2\frac{\Delta_F}{n+1}}
	+ \dots + e^{-\Delta_F}.
\eea
The MFPT is dominated by the $(n+2)$ term, while the SSTT is dominated by a constant (explicitly, by the $\Gamma_A^{-1}$ rate constant) and
the $e^{-\frac{\Delta_F}{n+1}}$ term. 
Therefore, by performing both transient and steady state measurements
one can experimentally determine both the number of sites $n$ and the potential drop $\Delta_F$.

{\it Positive potential bias, $E_D<E_A$.}
When the potential energy linearly increases in the direction of the trap,  
the rate constants satisfy $\kr{no} = e^{-\frac{\Delta_F}{n+1}}\text{, } 
\kl{no} = 1$. Note that we define the gap by its magnitude, $\Delta_F=|E_A-E_D|$.
The MFPT and the SSTT are
\bea 
\tau_m = 1 + (n+2)e^{\frac{\Delta_F}{n+1}} + (n+1) e^{2\frac{\Delta_F}{n+1}} + \dots + 2 e^{\Delta_F} 
\nonumber\\
\eea
and
\bea 
\tau_{ss} 
= e^{\frac{\Delta_F}{n+1}} + e^{2\frac{\Delta_F}{n+1}} + \dots + 2e^{\Delta_F}.
\eea
As $\Delta_F$ grows, both mean first-passage time and steady state transfer time approach 
$2e^{\Delta_F}$. This agreement is expected: when the forward energy gap is large, the 
steady state populations on  intermediate sites become small, and the timescales agree.

\subsection{Example III: Co-polymer motifs}
	
We consider polymers with two monomers, $B$ and $b$ of high energy ($E_B$) and low energy ($E_b$), 
with the scaled energy difference ${\Delta_P}=E_B-E_b$. 
The polymer may be alternating or stacked, e.g. with sequences BbBbBb or BBBbbb, respectively. 
For simplicity, the donor (acceptor) is placed in resonance with the first (last) site.
Examples of examined co-polymers are presented in Fig.\@ \ref{Figs2}.
In what follows, we study the dependence of the MFPT and the SSTT on $n$, 
the number of intermediate sites, the parity of $n$, 
and ${\Delta_P}$, the energy difference between monomers. 
Trends also depend on whether the donor state is coupled to the high or low energy monomer.
Our results are organized in Tables I and II.

\subsubsection{Alternating Polymers} 
Focusing first on the MFPT, we observe the following, see Table I:
(i) Polymers that end on the high energy monomer (sequences A2 and A3, such as DbBbBbBA or  DBbBbBA, respectively)
show the functional form $e^{{\Delta_P}}$, but not $e^{-{\Delta_P}}$. 
This is because the residence times for low energy monomers
increase with $e^{{\Delta_P}}$ while the residence time of the B monomer here does not depend on ${\Delta_P}$.
In contrast, polymers that end on the lower strand (DBbBbBbA or DbBbBbA) have both $e^{{\Delta_P}}$ and 
$e^{-{\Delta_P}}$ terms. In this case, high energy monomer have residence times that 
depend on $e^{-{\Delta_P}}$ while low-energy monomer contribute the residence time $e^{{\Delta_P}}$.  
(ii) For all four types of alternating chains, the MFPT scales as $n^2 e^{{\Delta_P}}$. In contrast, the SSTT 
shows two different scaling laws, depending on whether the donor is in resonance with B or b.
In the former case, $\tau_{ss}\propto n$, while in the latter case, the barrier energy shows up
and  $\tau_{ss}\propto ne^{{\Delta_P}}$.

\label{table1}
\begin{center}
\vspace{4mm}
Table I: Alternating polymer motifs and characteristic transfer times (leading terms in bold)\\
\begin{tabular}{|l|l|l|l|}
\hline
\,\,\,	&	{sequence} &{ \textbf{MFPT} } & {\textbf{SSTT}} \\ 
\hline 
A1. (even)\,\,     &  { D(Bb)$_m$A} \,\,    &  {${\bf e^{{\Delta_P}}}\left( {\bf \frac 14 n^2} - \frac 12 n \right) + \left(n + 2\right) e^{-{\Delta_P}}   + \frac 14 n^2 + 2n + 1$}  
& {${\bf n}+ 2e^{-{\Delta_P}} $} \\
 & $m=n/2$& & \\
 \hline
A2. (even)\,\, &  {D(bB)$_{m}$A} \,\,\, 		& 
{${\bf e^{{\Delta_P}}}\left( {\bf \frac 14 n^2} + 2n + 1 \right)  + \frac 14 n^2 + \frac 12 n + 2$} & 
{${\bf e^{\Delta_P}}({\bf n} + 1) + 1$ }\\
 & $m=n/2$& & \\
 \hline
A3. (odd)\,\,  	&		{D(Bb)$_{m}$BA } \,\,\, &
			{${\bf e^{{\Delta_P}}}\left( {\bf \frac 14 n^2} + \frac 12 n - \frac 34 \right)  + \frac 14 n^2 + 2 n + \frac{15}{4}$}   & {${\bf n}+2$} \\
& $m=(n-1)/2$& & \\
\hline
A4. (odd)\,\,	&	    	
{D(bB)$_{m}$bA} \,\,\,  	& 
{${\bf e^{{\Delta_P}}} \left({\bf \frac 14 n^2} + n - \frac 54 \right) + (n-1)e^{-{\Delta_P}}   + \frac 14 n^2 + \frac 12 n + \frac{21}{4} $}  & {${\bf e^{{\Delta_P}}}({\bf n} - 1)  + 3 $} \\
 &$m=(n-1)/2$ & & \\
	\hline

\end{tabular} 
\end{center}

\subsubsection{Stacked polymers}
In this model, the co-polymer is made of two halves of different content, which could 
correspond for example to different base-pairings in DNA molecules, see Ref. \cite{DNAS1,DNAS2}.
In this case, all rate constants $\kl{no}, \kr{no}$ are set equal to 1, 
besides a single rate where the monomers switch. 
For the mean first-passage time, this splits the sum of residence times into two groups: 
residence times corresponding to sites 
after the change in energy, which do not depend on ${\Delta_P}$, 
and times corresponding to sites before the change in energy, which do manifest a ${\Delta_P}$-dependence. 
The results are quite natural, see Table II:
If we switch from high energy sites towards low energy sites as we move towards the acceptor, we do not pay an energetic price
and $\tau_m \propto n^2$, $\tau_{ss}\propto n$.
In the opposite case, these times are increased by the $e^{{\Delta_P}}$ factor.
%

\begin{center}
\vspace{4mm}
Table II: Stacked Polymer motifs and characteristic transfer times (leading terms in bold)\\
\label{table1}
\begin{tabular}{|l|l|l|l|}
\hline
\,\,\,\,	&	{sequence} &{ \textbf{MFPT} } & {\textbf{SSTT}} \\ 
\hline 
S1. (even)\,  	&		{DB$_{m}$b$_m$A } \,\,\,  & $ { \bf \frac 14 n^2 }+ \frac 32 n + 2 + \left( \frac 14 n^2 + n + 1 \right) e^{-{\Delta_P}} $  & 
$ {\bf \frac 12 n} + 1 + \left( \frac 12 n + 1 \right) e^{-{\Delta_P}}  $\\
& $m=n/2$ & & \\
\hline
S2. (even)\,  	&		{Db$_{m}$B$_m$A } \,\,\, &  ${\bf e^{\Delta_P} } \left( {\bf \frac 14 n^2} + \frac 32 n + 2 \right) + \frac 14 n^2 + n + 1$ & ${\bf e^{\Delta_P}}\left( {\bf \frac 12 n} + 2 \right) + \frac 12 n $ \\
& $m=n/2$& & \\
\hline
S3. (odd)\,  	& {DB$_{m}$b$_{m+1}$A	} \,\,\, 	&   ${ \bf \frac 14 n^2} + \frac 32 n + \frac 94 + \left( \frac 14 n^2 + n + \frac 34 \right) e^{-{\Delta_P}} $ & 
$ { \bf \frac 12 n }+ \frac 12 + \left( \frac 12 n + \frac 32 \right)e^{-{\Delta_P}} $\\
&$m=(n-1)/2$ && \\
\hline
S4. (odd)\,  	&		{Db$_{m}$B$_{m+1}$A } \,\,\, &  ${\bf e^{{\Delta_P}}}\left( {\bf \frac 14 n^2} + \frac 32 n + \frac 54 \right)  + \frac 14 n^2 + n + \frac 74 $ & 
${\bf e^{{\Delta_P}}}\left({\bf \frac 12 n} + \frac 52 \right)  + \frac 12 n - \frac 12 $\\
& $m=(n-1)/2$ && \\
	\hline
\end{tabular} \\
\end{center}

Overall, we find that:
(i) The MFTT and SSTT of alternating configurations 
are typically longer than for a stacked one.  
It is interesting to note that higher 
resistances were indeed measured for alternating DNA configurations, 
relative to stacked sequences \cite{DNAS1,DNAS2}, 
in line with our calculations. 
Nevertheless, since coherent quantum effects are not included in this work, 
this agreement may be incidental. 
(ii) If $n$ and ${\Delta_P}$ are large enough such that only the leading term matters,
it is generally possible to determine both ${\Delta_P}$ and $n$ by jointly studying  the
steady state lifetime and the mean first-passage time,  assuming we know the nature of
the polymer (from the configurations S1-S4 and A1-A4), but not necessarily its length or energetics.
First, we note that $\tau_m / \tau_{ss} \propto n$ in all cases, besides A1 and A3. 
In these special cases, the dominating term of steady state lifetime is itself proportional to $n$.
As well, all co-polymers besides S1 and S3, 
show  $\frac 14 n^2e^{{\Delta_P}}$ as the leading term in the mean first-passage time.
In all these configurations, ${\Delta_P}$ can be readily determined once we find $n$.


\vspace{10mm}

\begin{figure}[htbp]
\includegraphics[scale=0.3]{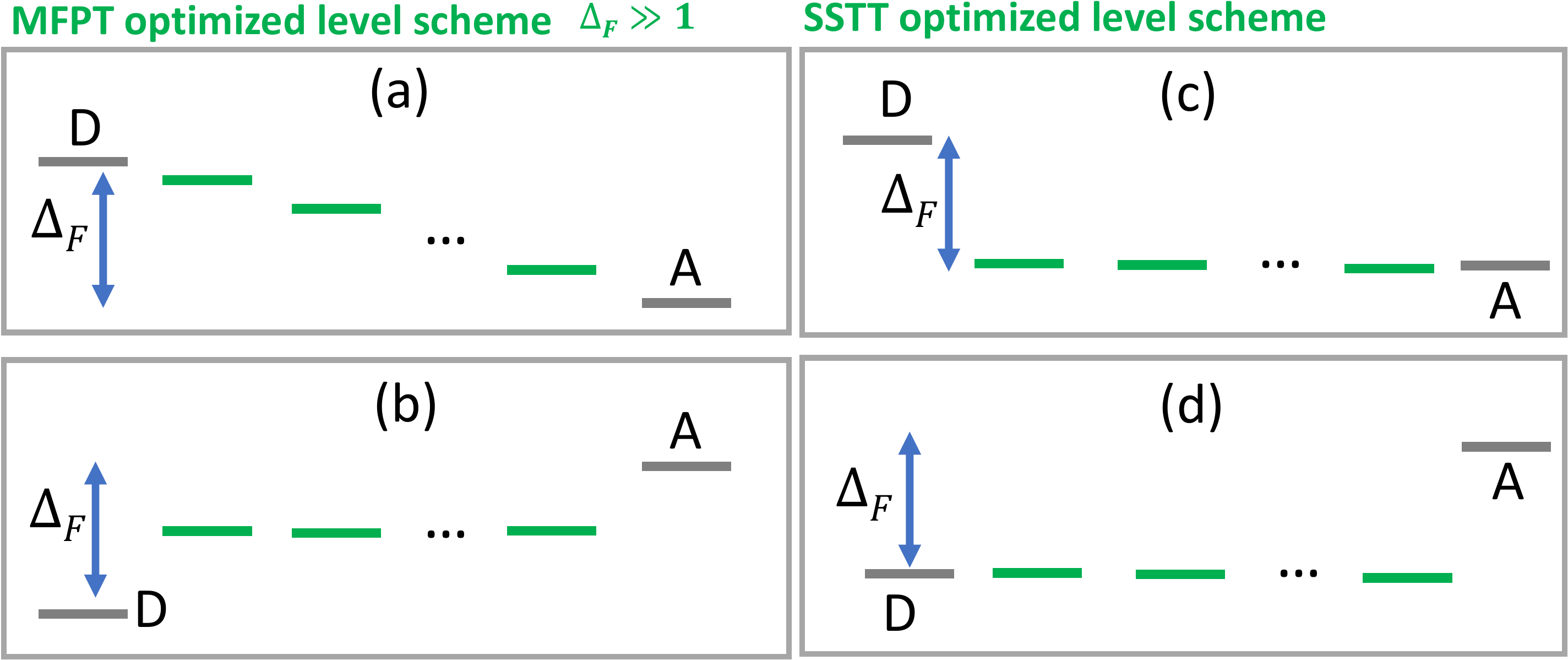}  
\caption{Illustrations of energy profiles that minimize the (a)-(b) MFPT  and (c)-(d) SSTT
under forward or reversed biases.
The energy gap between the D and A sites is fixed. 
The position of the intermediate levels is determined by minimizing the transfer time.
Under low bias (not shown), the energy profile minimizing the MFPT is nonmonotonic, 
and it generates an internal strong field  as we show in Figs. 
\ref{fig-negative-energy-profile} and \ref{fig-positive-energy-profile}.
}
\label{fig-opt}
\end{figure}

\section{Optimization of the energy profile} 

Achieving fast rate of charge transport through molecules is desirable for many applications. Experiments
manifest an enhancement of hole transport in DNA by modifying the injection site, the sequence,
and the chemical environment. These modifications alter the molecular energy profile 
thus the driving force for charge migration \cite{speed, lipid,Grozema}.

In this Section, we ask the following question: Assuming a fixed donor and acceptor sites energies, 
what is the optimal energy profile for the intermediate $n$ sites, 
so as to minimize the transfer time---thus enhance the speed of 
transfer? We explore this question analytically and numerically for positive and negative biases. 
Central observations are that the MFPT and the SSTT are minimized under different principles, and
that the profile generally deviates from linearity and monotonicity.
In Fig. \ref{fig-opt} we exemplify optimized setups under large bias, derived in the following discussion.

\subsection{Negative bias, $E_D>E_A$}
\subsubsection{MFPT}
Let us first consider the negative bias case, with a total energy gap of $\Delta_F=|E_D-E_A|$ 
between the donor and acceptor.
As a constraint, we demand that the energies of the sequence of $n$ internal sites are non-increasing.
We assign a (temperature scaled) energy $E_i$ to site $i$. The transition rate constant from site $i$
to $i-1$ is set at $\kl{i-1}=e^{E_{i} - E_{i-1}}$ if $E_i < E_{i-1}$ and $1$ otherwise.
For simplicity, the trapping rate constant is taken as $\Gamma_A = 1$.
%
Since the energies do not increase going towards the trap, all $\kr{i} = 1$, and the MFPT follows Eq.\@
(\ref{eq:descendmftt}). The overall energy gap satisfies the constraint 
\bea 
\prod_{i=0}^{n} \kl{i} = e^{-\Delta_F}. 
\eea
The optimal rates in Eq.\@ (\ref{eq:descendmftt}) fulfill the constraint and $n+1$ equations of the form
\begin{align}
\label{LMTmfpt} 
	\frac{\partial \tau_m}{\partial \kl{i}} =&
	1 + \kl{i-1} +  \kl{i+1}+ 
	\kl{i-2}\kl{i-1} + \kl{i-1}\kl{i+1} +  \kl{i+1}\kl{i+2}  \dots  
	\nonumber\\
	+& \kl{0} \kl{1} \dots \kl{i-1}\kl{i+1} \dots \kl{n} 
	\nonumber\\
	=& \lambda \kl{0} \kl{1} \dots \kl{i-1}\kl{i+1} \dots \kl{n} 
\end{align}
where the constant $\lambda$ is the Lagrange multiplier;   $\kl{j} = 0$ if $j<0$. 
As an example, let us consider the case with two intermediate sites, $n = 2$.
The MFPT is 
\bea
\tau_m= 4+ \kl{0}+\kl{1}+\kl{2} + \kl{0}\kl{1} + \kl{1}\kl{2} + \kl{0}\kl{1}\kl{2},
\eea
and the following four equations should be jointly solved 
\begin{subequations}
 \begin{align} 
                1 + \kl{1} + \kl{1}\kl{2} &= \lambda \kl{1}\kl{2} \\
                1 + \kl{0} + \kl{2} + \kl{0}\kl{2} &= \lambda \kl{0}\kl{2} 
\label{eq:b}
\\
                1 + \kl{1} + \kl{0}\kl{1} &= \lambda \kl{0}\kl{1} \\
                \kl{0}\kl{1}\kl{2} &= e^{-\Delta_F} 
\end{align}
\end{subequations}
From symmetry, note that the first and last transition rates are identical, $\kl{2}=\kl{0}$.
In general,  since Eq.\@  (\ref{LMTmfpt}) is symmetric with respect to replacing each $\kl{i}$ with $\kl{n-i}$,
pairs of rate constants of equal distance from the first and last sites will be identical.
For $n=2$, the minimization problem involves solving 
a third-degree polynomial in $\sqrt{\lambda}$,
\bea 
0 = \lambda \sqrt{\lambda} - 2 \lambda + \sqrt{\lambda} - e^{\Delta_F}.
\label{eq:LMTpolynomial}
\eea
where
\bea \kl{0} = \kl{2} = \frac{1}{\sqrt{\lambda} - 1}, \, \kl{1} = \frac{1}{\sqrt{\lambda}}.
\eea
We can also organize Eq. (\ref{eq:LMTpolynomial})  as $x(x-1)^2 = e^{\Delta_F }$ with  $x=\sqrt{\lambda}$.
As expected, when $\Delta_F$ becomes large, $x$ grows and approaches $x\sim e^{\Delta_F/3}$,
which corresponds to constant spacings (linear profile).
Optimization of the model with $n$ intermediate sites 
requires solving a similar $n+1$-degree polynomial.

%
\begin{figure} [htbp]
\includegraphics[scale=.4]{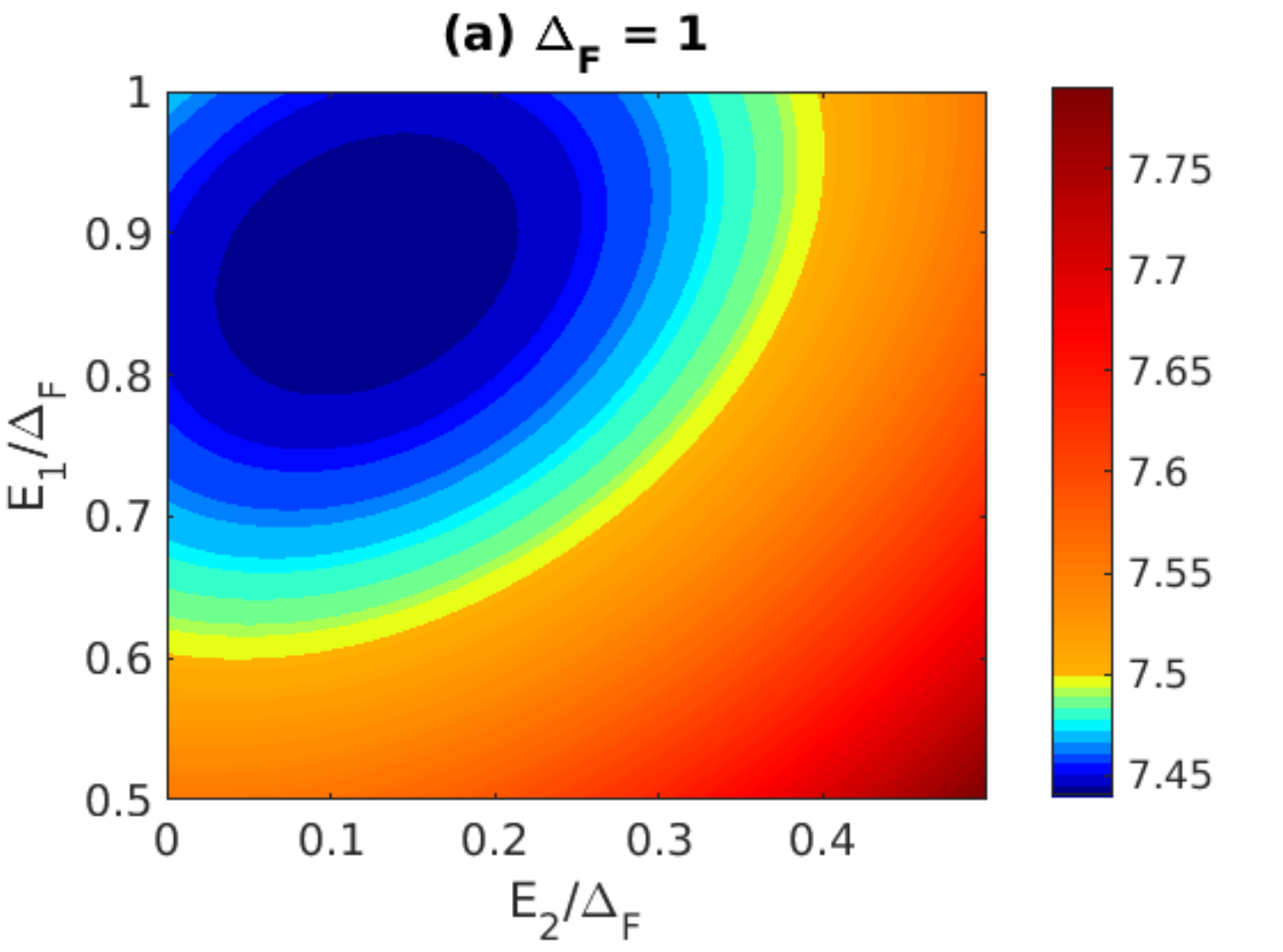}    
\includegraphics[scale=.4]{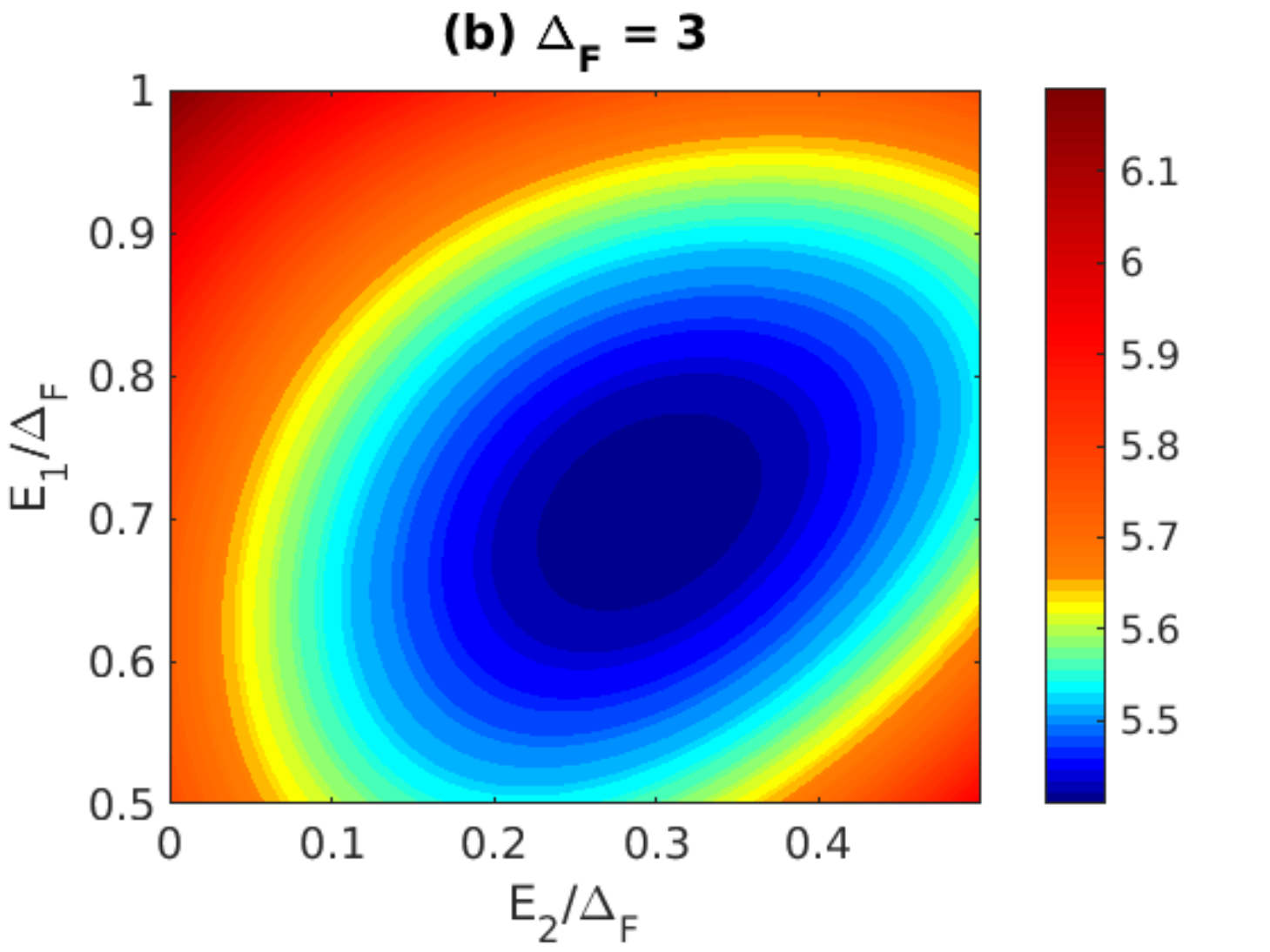} 
\includegraphics[scale=.4]{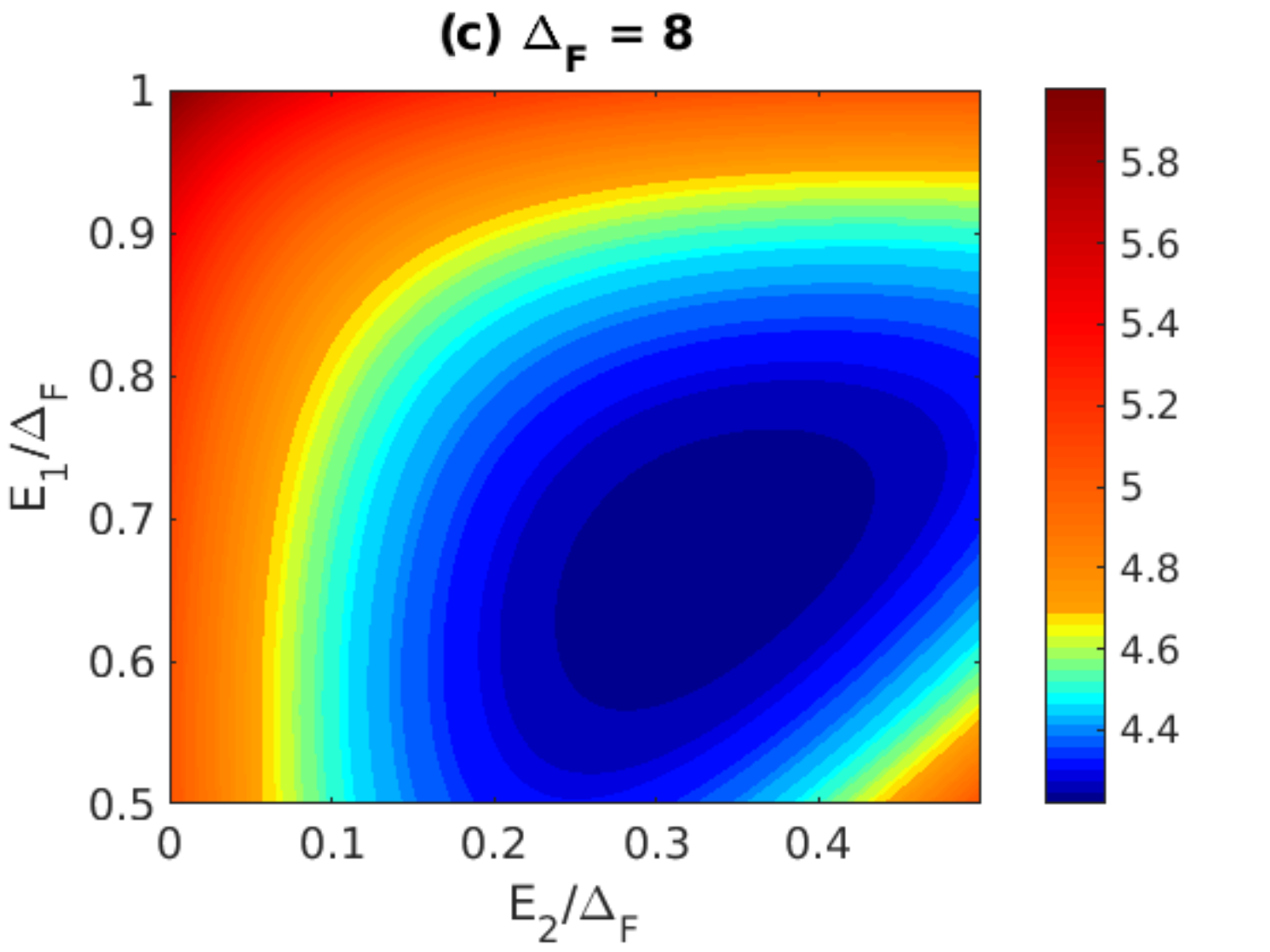}
\caption{Simulations of the mean first-passage 
time for configurations with two internal sites (D, 1, 2, A) for negative bias, $E_D>E_A$. 
As $\Delta_F$ increases, the optimal configuration (minimum MFPT)
approaches the profile  $(\frac{2}{3}\Delta_F,  \frac{1}{3}\Delta_F)$.}
\label{fig-2d-negative-bias}
\end{figure}

\begin{figure} [htbp]
\includegraphics[scale=.7]{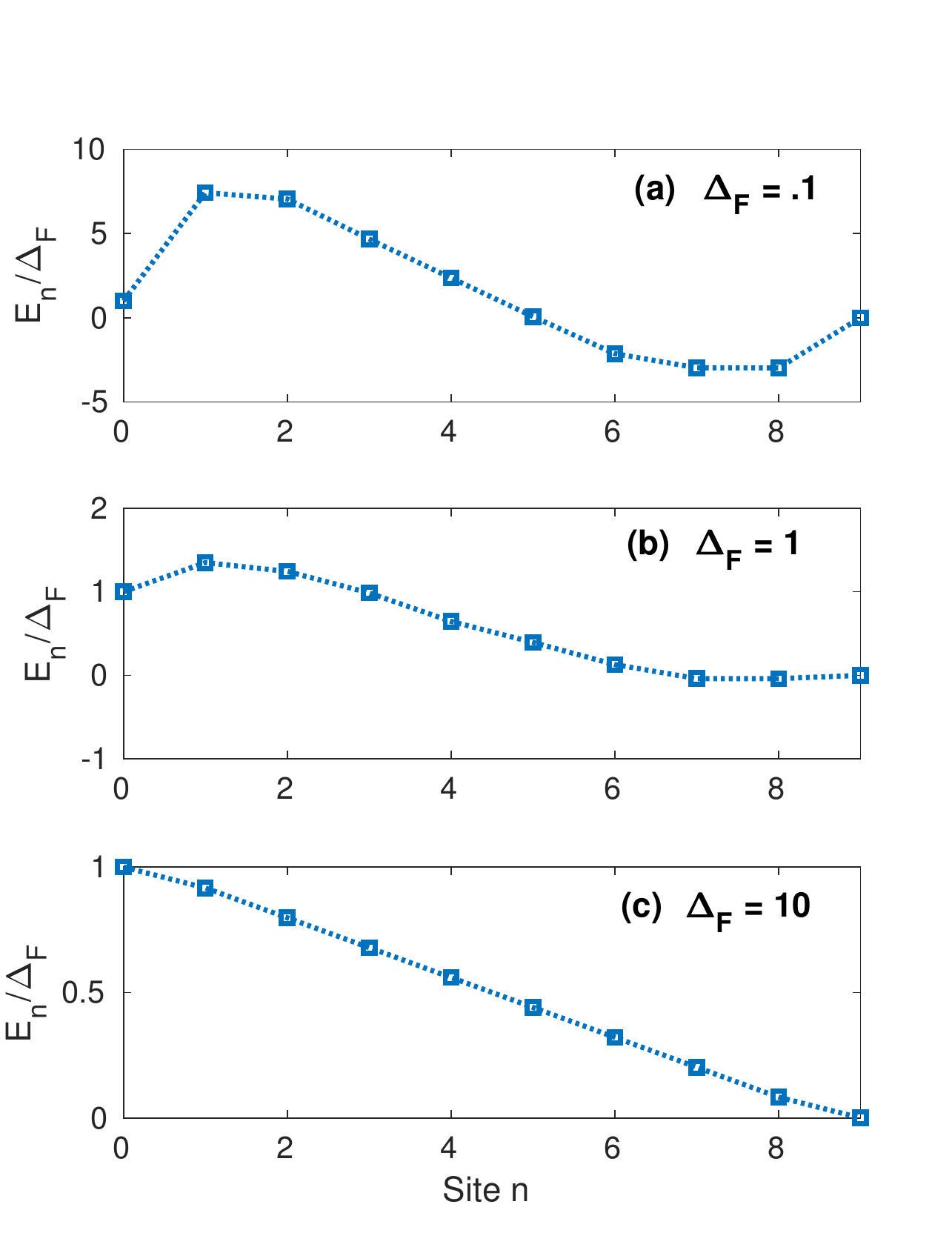} 
\caption{Negative bias, $E_D>E_A$. 
Energy profile obtained numerically using gradient descent 
by minimizing the MFPT for systems with eight internal sites (D, 1, \dots 8, A). }
\label{fig-negative-energy-profile}
\end{figure}

The analytic calculation above constrains the energies to be non-increasing. 
This constraint is not enforced in simulations that in fact reveal situations in which
the transfer times are minimized with nonmonotonic profiles. 
In Figures \@ \ref{fig-2d-negative-bias}-\ref{fig-negative-energy-profile} 
we demonstrate such calculations using gradient descent.
We find that when $n$ is small ($n=2$) and $\Delta_F$ is large, 
starting from the D, the energies descend slowly (small local bias), then more quickly towards the
center (large local bias), and again slowly at the end of the chain.
In contrast, when $n$ is large and $\Delta_F$ small, the first internal site is placed {\it above} the donor state.
The rest of the sites descend linearly, reaching below the acceptor state, see Fig. \ref{fig-negative-energy-profile}(a)-(b).
This result can be rationalized as follows: The system pays the cost of a
large energy jump in the first transition (slow-down of transfer), yet this early barrier 
prevents the carrier from going backwards towards the donor from the internal sites.
Nevertheless, if we force the internal sites to take values only in between the donor and acceptor levels,
we obtain profiles that are similar to the small-$n$ large-$\Delta_F$ case.

Non-monotonic internal profiles allow fast transfer when the external bias $\Delta_F$ is small relative to the 
temperature.
In such low-field cases, the levels organize to create a large internal field,  opposing thermal
fluctuations and facilitating fast transport.
This principle could be used in the design of biomolecules that  support  fast charge migration \cite{Grozema}.

\subsubsection{SSTT}
Continuing with the same model, with a non-increasing profile, we turn to the steady state transfer time, 
\bea
\label{descendkss}
 \tau_{ss} = \kl{0} + \kl{0}\kl{1} + \kl{0} \kl{1} \kl{2} + \dots + \prod_0^{n} \kl{i}.
\eea
This expression is minimized when $\kl{0} = e^{-\Delta_F}$, $\kl{i} = 1$ for
 $i \neq 0$, resulting in the  minimum steady state transfer time of $(n+1)e^{-\Delta_F}$.
In other words, the fastest process occurs when all the internal sites are aligned with the acceptor, 
i.e., the energy jump occurs immediately, from the donor to the first site, $n=1$.

%

\subsection{Positive bias, $E_D<E_A$}
\subsubsection{MFPT}
\begin{figure} [htbp]
\includegraphics[scale=.4]{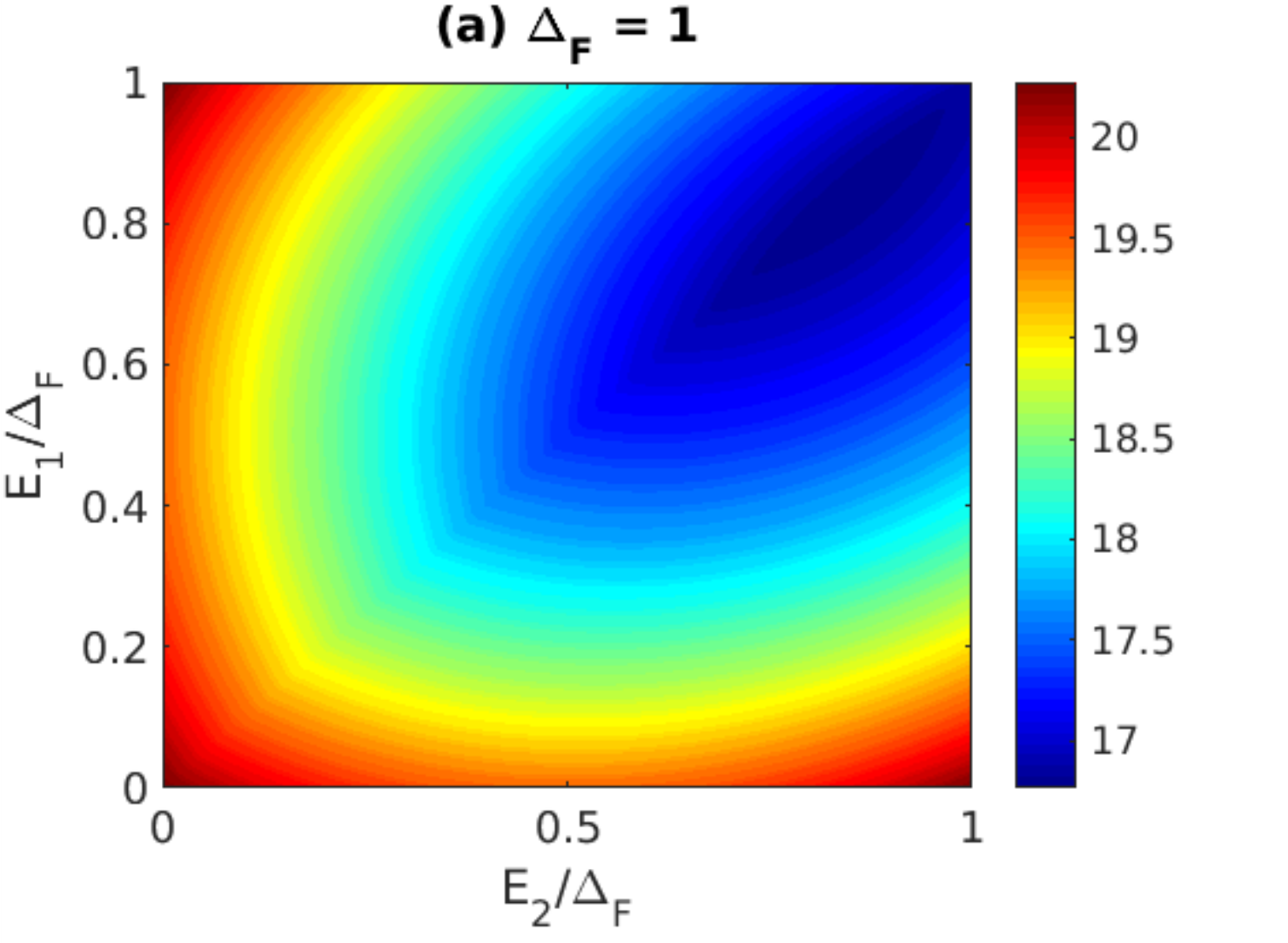}  
\includegraphics[scale=.4]{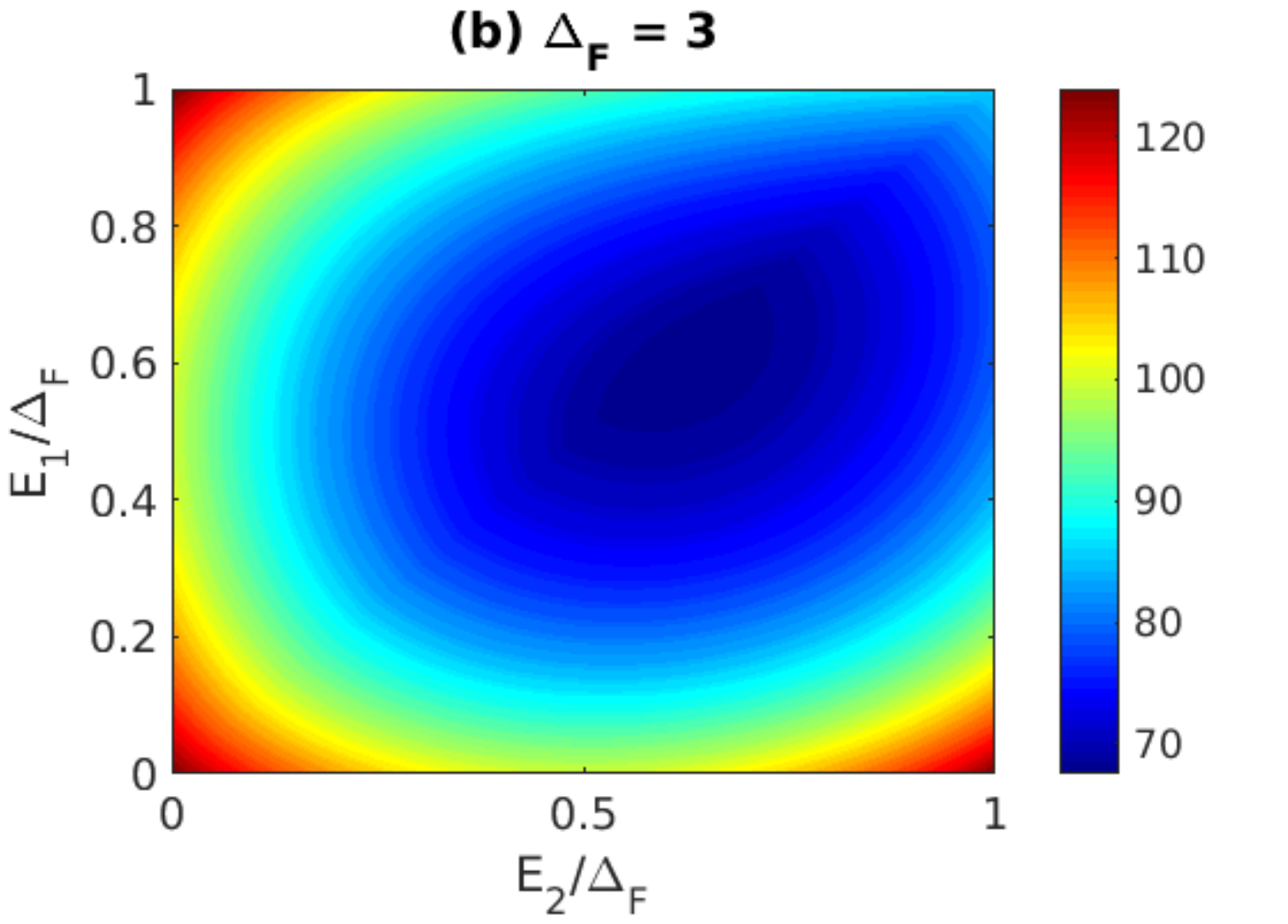} 
\includegraphics[scale=.4]{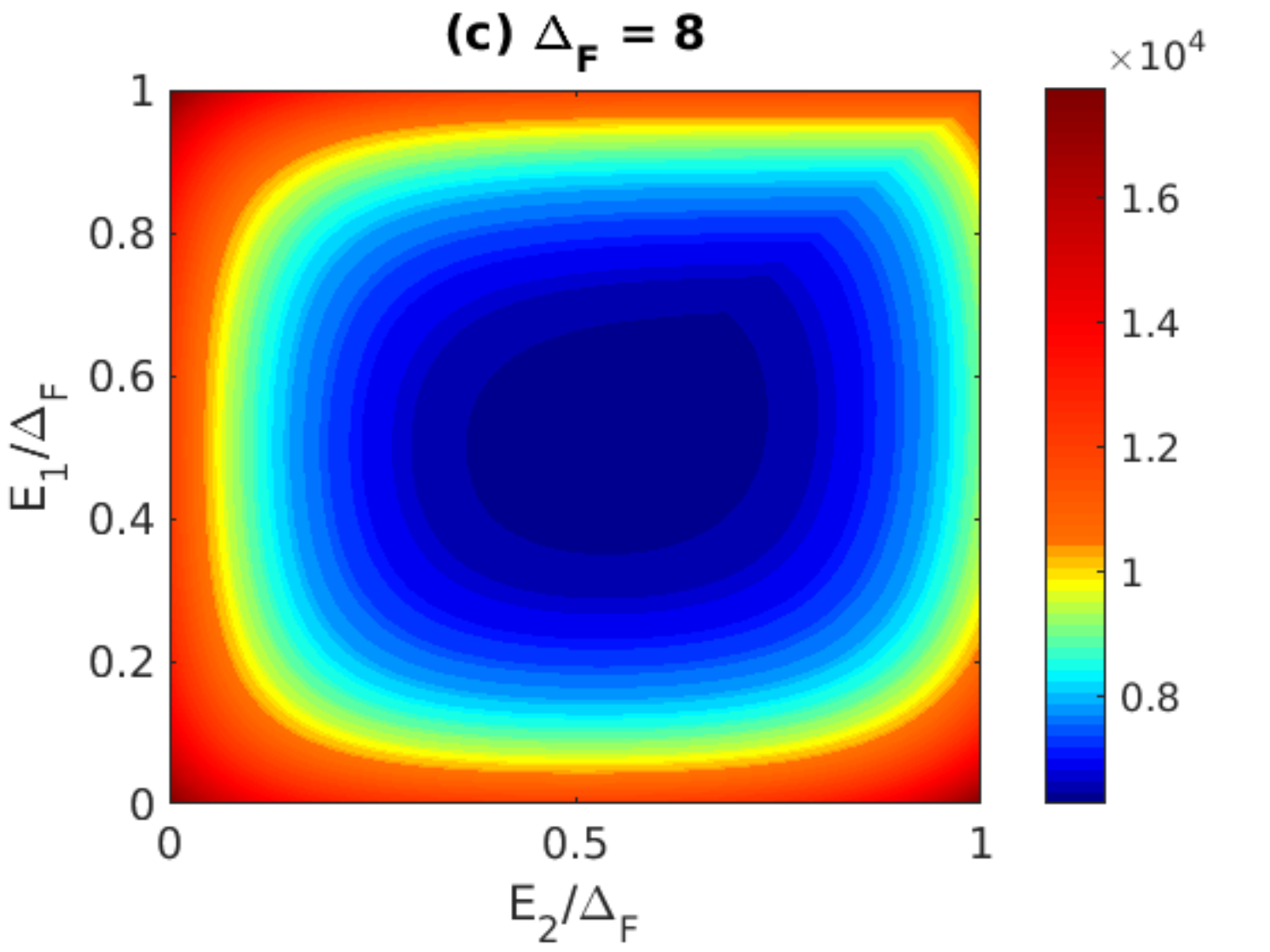}  
\caption{Simulations of the 
mean first-passage time for a two-state system with $E_D<E_A$.
\label{fig-2d-positive-bias}
As the gap $\Delta_F$ increases, the optimal configuration is of 
two states placed together at around $\Delta_F/2$.}
\end{figure}

\begin{figure} [htbp]
\includegraphics[scale=.7]{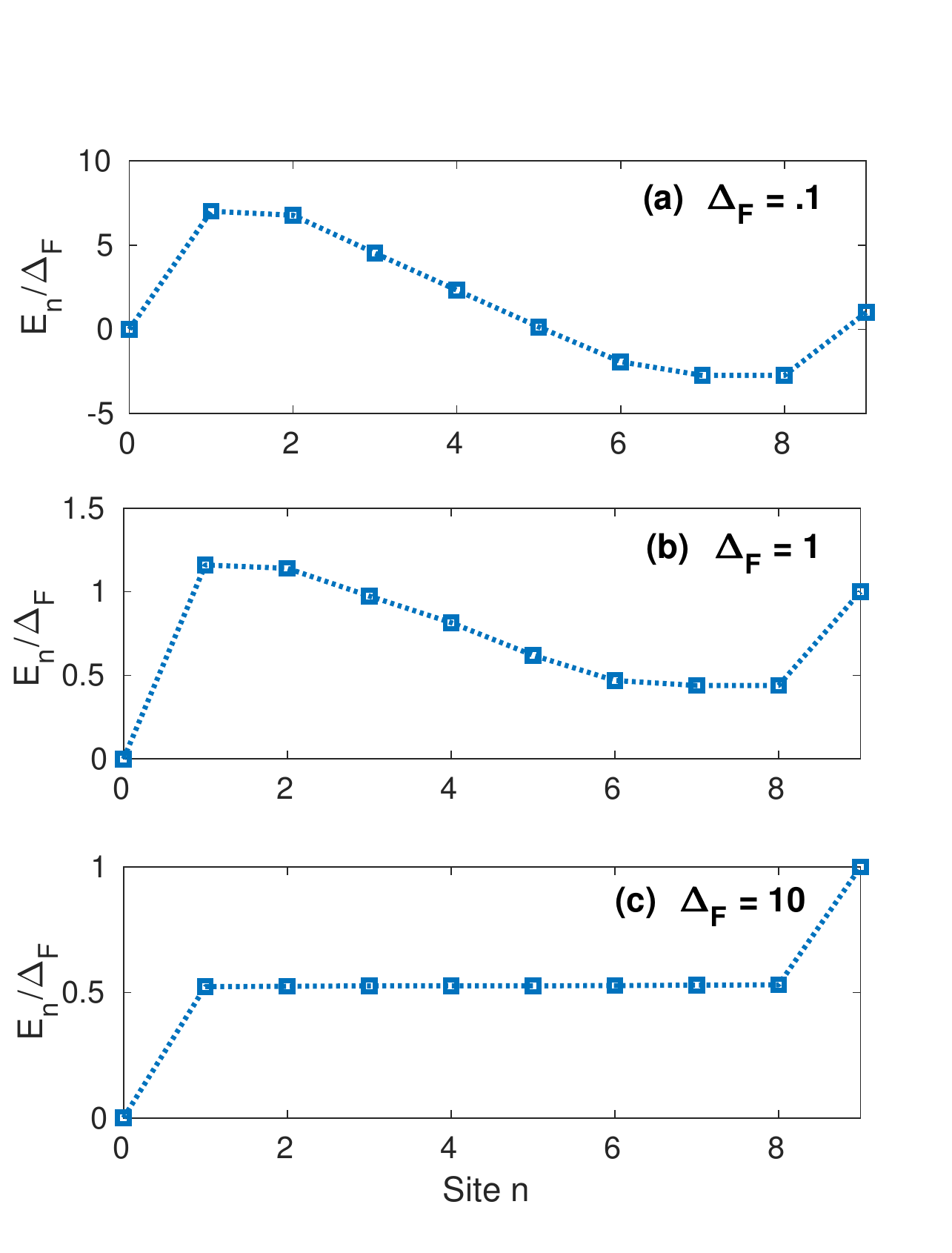}
\caption{Energy profile obtained numerically by minimizing the 
MFPT for a system with eight intermediate sites under a positive bias $E_D<E_A$.
At small $\Delta_F$, the profiles under positive and negative biases look identical,
compare panel (a) to Fig. \ref{fig-positive-energy-profile}(a).
}
\label{fig-positive-energy-profile}
\end{figure}

We now consider a low-energy donor, 
$n$ non-decreasing energy levels, and a high energy acceptor site.  
We define the gap as $\Delta_F=|E_D-E_A|$.
This arrangement implies that $\kl{no} = 1$, while the forward rates are to be optimized. 
The MFPT is
\begin{align}
\tau_m &= 
\left(\frac 1 \Gamma_A + \frac{1}{\kr{n+1}} + \dots + \frac{1}{\kr{1}} \right)  
\nonumber\\
&+ \left(\frac{1}{\Gamma_A \kr{n+1}} + \frac{1}{\kr{n+1}\kr{n}} + \dots + 
\frac{1}{\kr{2}\kr{1}} \right) 
 \nonumber\\
&+ \dots + 
\frac{1}{\Gamma_A \prod_{i=1}^{n+1} \kr{i}
}.
\end{align}
For simplicity, taking $\Gamma_A = 1$, we get
\begin{align}
\tau_m &= 
\left(\frac{2}{\kr{n+1}} + \frac{1}{\kr{n}} \dots + \frac{1}{\kr{1}} \right)  
\nonumber\\
&+
\left(\frac{2}{\kr{n+1}\kr{n}} + \frac{2}{\kr{n}{\kr{n-1}}} +  \dots +
\frac{1}{\kr{2}\kr{1}} \right)  
\nonumber\\ 
&
+ \dots + 
\frac{2}{\prod_{i=1}^{n+1} \kr{i}}.
\end{align}
Minimizing the MFPT subjected to the constraint $\prod_{i=1}^{n+1} \kr{i} = e^{-\Delta_F}$,
the equations have no convenient symmetries. 
For example, $n = 1$ yields
\bea
\tau_m=1 + \frac{1}{\kr{1}} + \frac{2}{\kr{2}} +\frac{2}{\kr{1}\kr{2}}.
\eea
While this equation can be readily minimized with the constraint, we
proceed with the general formalism of Lagrange multipliers
and solve the system,
\begin{subequations}
\begin{align}
-\frac{1}{\kr{1}^2} - \frac{2}{\kr{1}^2\kr{2}} &= \lambda \kr{2}\\
-\frac{2}{\kr{2}^2} - \frac{2}{\kr{1}\kr{2}^2} &= \lambda \kr{1}\\
\kr{1}\kr{2} &= e^{-\Delta_F}.
\label{eq:c}
\end{align}
\end{subequations}
This problem can be solved by  noting that
\bea 
-\frac{1}{\kr{1}^2\kr{2}} - \frac{2}{\kr{1}^2\kr{2}^2} 
= \lambda = 
-\frac{2}{\kr{1}\kr{2}^2} - \frac{2}{\kr{1}^2\kr{2}^2},
\eea
which implies that
$\kr{2} = 2\kr{1}$, or explicitly
\bea 
\kr{1} = e^{\frac 12 \left( -\ln 2 - \Delta_F \right)}, \quad 
\kr{2} = e^{\frac 12 \left( \ln 2 - \Delta_F \right)}.
\eea
Figures \@  \ref{fig-2d-positive-bias}-\ref{fig-positive-energy-profile} present simulation results, where
we allowed the internal potential profile to become nonmonotonic (unlike the analytic derivation).
We find that for large $n$ and large $\Delta_F$,
the internal sites cluster at a level slightly above the midgap
between the donor and acceptor. 
In contrast,  for very small $\Delta_F$ the energy profile first increases,
thus artificially creating a large internal field, which facilities fast transfer. 
In fact, when $\Delta_F$ is small the results for positive and negative biases are
very similar, compare Fig. \ref{fig-negative-energy-profile}(a) to Fig. \ref{fig-positive-energy-profile}(a).

\subsubsection{SSTT}
We calculate the SSTT using Eq. (\ref{eq:ssTTc}), assuming now a non-decreasing profile,
\bea  
\tau_{ss} = \left[\kr{1}^{-1}  + \left(\kr{1} \kr{2}\right)^{-1} + \dots +
\prod_{i=1}^{n+1} \kr{i}^{-1}. \right]
\label{eq:ssfincrease} 
\eea
The minimum MFPT, subject to the constraint $\prod_{i=1}^{n+1} =
e^{-\Delta_F}$, can be intuitively obtained: 
each term in Eq.\@ (\ref{eq:ssfincrease}) 
is greater  or equal of the preceding term. Therefore, $\tau_{ss}$ is minimized
once we take $\kr{n+1} =
e^{-\Delta_F}$ and $\kr{i} = 1$ for all $i \neq n+1$. 
This setup is illustrated in Fig. \ref{fig-opt}(d): besides the last site, all sites align with the donor state.

\section{Summary}

In summary, using kinetic equations we studied particle transfer in one-dimensional systems with different motifs:
linear, branched, uniform, biased, homogeneous and multi-component systems.
Central results of our work are:
(i) We derived an intuitive relationship between the MFPT and the SSTT.
(ii) We obtained closed-form expressions for the MFPT and the SSTT in nearest-neighbor 1D chains.
(iii) We exemplified our results on experimentally relevant setups, such as donor-bridge-acceptor systems, biased
chains, and stacked and alternating co-polymers, and discussed the physical information
that can be gained from the MFPT and the SSTT.
(iv) We minimized the transit time through chains
by optimizing the energy profile. 
Here, we found that the MFPT and the SSTT were minimized under fundamentally different design rules.
Most strikingly, under shallow external potentials of the order of the thermal energy, $\Delta_F\leq 1$,
the system achieved fast transit time if the levels were set so as to create a strong internal 
field.

This work clarifies on the relationship between transient and steady state measures in classical kinetic networks.
It is worthwhile to mention studies of the MFPT in complex scale-invariant media \cite{klafter}
and in the context of enzymatic chemical reactions in biochemistry \cite{Udo07,Cao11}.
Nevertheless, in these works the networks included loops, a motif that was not investigated in the present study.
It is also interesting to consider extensions of this work, particularly the enhancement of the transfer speed, 
to systems under thermal gradients \cite{NitzanH}.
The analysis of transfer processes in networks beyond 1D, and the role of quantum coherent
effects in corresponding quantum systems, will be the focus of future works. 

\begin{acknowledgments} 
DS acknowledges the Natural Sciences and Engineering Research Council (NSERC) of Canada Discovery Grant 
and the Canada Chair Program.
NK was supported by the NSERC Undergraduate Student Research Award (USRA)  program.
\end{acknowledgments} 


\renewcommand{\theequation}{A\arabic{equation}}
\setcounter{equation}{0}  
\section*{Appendix: Graphs with multiple acceptors}

We consider here a system with a single donor and multiple acceptors.
Each acceptor is connected to the trap site,
therefore we define the trapping mean first-passage time as in Eq. (\ref{eq:MFPTdef}).
The definition of the MFPT in terms of the
residence times, Eq. (\ref{eq:MFPT1}), holds even with multiple acceptors.

\begin{figure}[h]
\includegraphics{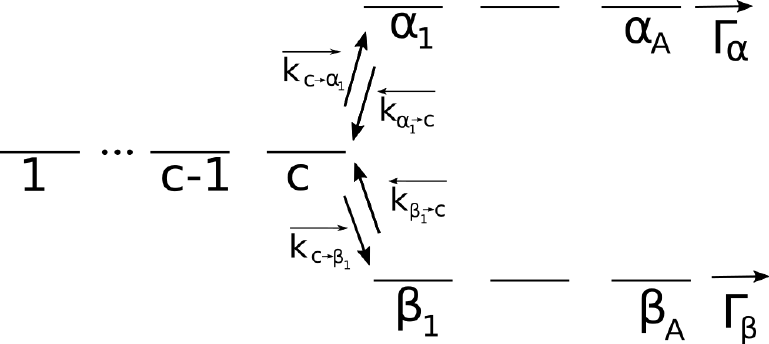}
\caption{A network with a single branching point $c$. 
The two branches are identified by $\alpha$ and $\beta$.}
\label{FigA}
\end{figure}

For simplicity, let us consider
a chain that is branched at point $c$ into two chains, $\alpha$ and $\beta$, with
two acceptors, which are identified by $\alpha_A$ and $\beta_A$,  see Fig. \ref{FigA}. 
We explore next the MFPT to be trapped leaving from either acceptors.
The normalization of the trap population  leads to 
$\int_0^{\infty}dt \left(\Gamma_{\alpha}p_{\alpha_A}(t) +  \Gamma_{\beta}p_{\beta_A}(t)\right)  =1$.
In the language of residence time this condition translates to
$\Gamma_{\alpha}r_{\alpha_A}+\Gamma_{\beta}r_{\beta_A}=1$.
Furthermore, it is useful to define the yield to each acceptor as 
\bea \Phi_{l} \equiv r_{l_A} \Gamma_{l}, \,\,\,\ l=\alpha,\beta \eea
with the total yield $\sum_l{\Phi_l}=1$.

Based on the discussion of Sec. \ref{3MFPT}, the MFPT
for a graph with no loops is determined by a flux condition,
%
\bea \kr{no} r_i - \kl{no} r_{i+1} = \Phi, 
\eea
with $\Phi$ the yield on the branch considered.
To find the trapping MFPT, we need to resolve the residence times at every site. 
Particularly, if we find
the yields (proportional to residence times) at the acceptors, 
we can use the flux condition and moving backwards generate the residence times at every site.
We now describe how to find the yields at each acceptor for a system with a single branching point, Fig. \ref{FigA}.
The method can be furthermore generalized to describe a more complex graph, with additional branches.

In the model sketched in Fig. \ref{FigA},  site $c$ is connected to three sites, 
$c-1$, $\alpha_1$, and $\beta_1$, where
$\alpha_1$ and $\beta_1$ are the first sites in chains that end in acceptors
$\alpha_A$ and $\beta_A$, respectively. These sites have respective rates to the trap, $\Gamma_\alpha$, $\Gamma_\beta$.
Once we determine the yields
 $\Phi_{\alpha}$ and  $\Phi_{\beta}$, we can proceed and find the residence times at all sites 
by the flux condition.

To find the yield $\Phi_{\alpha}$, we consider a modified system composed of 
the sites $\alpha_1 \dots \alpha_A$,
with $\alpha_1$ as an additional acceptor,
 with an irreversible rate $\kl{\alpha_1 \to c}$. For this modified system, we define 
the yield through the `true' acceptor as $\Phi^*_{\alpha}$, 
which is the probability that, given a particle arriving at $\alpha_1$, it will exit at
$\alpha_A$ and not at $\alpha_1$. In the original model, this yield represents the 
probability that a particle
coming to $\alpha_1$ will leave by $\alpha_A$ without returning to $c$.

The yield $\Phi^*_{\alpha}$ can be readily obtained analytically, since
the matrix corresponding to the new
system is tridiagonal. The solution is found using the Thomas algorithm;
the rate matrix has the same form as in Eq. (\ref{eq:arbratematrix}) except the top left element 
combines transitions from $\alpha_1$ to both $\alpha_2$ and the `trap' $c$,
 $-\kr{\alpha_1\to\alpha_2}-\kl{\alpha_1\to c}$.
%
The yield obtained is
\bea
\Phi^*_{\alpha} 
= \left( 1
+ \kl{\alpha_1 \to c} \left( \sum_{i=1}^{n-1} \frac{\prod_{j=1}^i\kl{\alpha_{j+1}\to \alpha_j}}{\prod_{j=1}^{i+1} \kr{\alpha_{j-1} \to \alpha_j}}
+ \frac{\prod_{j=1}^{n}\kl{\alpha_{j+1}\to \alpha_j}}{\Gamma_\alpha\prod_{j=1}^{n+1}\kr{\alpha_{j-1} \to \alpha_j}} \right)\right)^{-1},
\eea
%
where $n$ is the number of non-acceptor states, and site $\alpha_{n+1} \equiv \alpha_{A}$.
As expected, if $\kl{\alpha_1 \to c} = 0$ then $\Phi^*_{\alpha} = 1$;  increasing $\kl{no}$ relative to $\kr{no}$ reduces the yield. The yield $\Phi^*_{\beta}$ can be calculated in a similar way.

We now relate the true yields $\Phi_{\alpha,\beta}$ to the auxiliary ones $\Phi_{\alpha,\beta}^*$.
The probability that the particle exits at $\alpha_A$ after visiting $c$ only once
is
\bea 
\frac{\kr{c \to \alpha_1}}{\kr{c \to \alpha_1} + \kr{c \to \beta_1}} \Phi^*_{\alpha} 
\eea 
Here, the probability to enter $\alpha_1$ is multiplied by the chance
to leave from $\alpha_A$ without returning to $c$, given that the particle reaches $\alpha_1$.

Likewise, the probability that the particle exits from $\alpha_A$, leaving $c$ to the
right exactly twice is
\bea 
\left( \frac {\kr{c\to\alpha_1}\left(1-\Phi^*_{\alpha}\right) +
\kr{c\to\beta_1}\left(1-\Phi^*_{\beta}\right)}{\kr{c\to\alpha_1} + \kr{c\to\beta_1} }\right)
\frac{\kr{c \to \alpha_1}}{\kr{c \to \alpha_1} + \kr{c \to \beta_1}}  \Phi^*_{\alpha},
\eea 
where the new term in the product is the probability to return to $c$ after leaving to the right once.
Overall, the probability of leaving site $c$ $n$ times to the right is 
\bea 
\left( \frac {\kr{c\to\alpha_1}\left(1-\Phi^*_{\alpha}\right) +
\kr{c\to\beta_1}\left(1-\Phi^*_{\beta}\right)}{\kr{c\to\alpha_1} + \kr{c\to\beta_1} }\right)^{n-1}
\frac{\kr{c \to \alpha_1}}{\kr{c \to \alpha_1} + \kr{c \to \beta_1}}  \Phi^*_{\alpha}. 
\eea 
The total yield is a geometric series, which collapses to the simple result,
\bea
\label{eq:fluxfromlinflux}
\Phi_{\alpha} &=& \sum_{n=1}^\infty 
\left( \frac {\kr{c\to\alpha_1}\left(1-\Phi^*_{\alpha}\right) +
\kr{c\to\beta_1}\left(1-\Phi^*_{\beta}\right)}{\kr{c\to\alpha_1} + \kr{c\to\beta_1} }\right)^{n-1}
\frac{\kr{c \to \alpha_1}}{\kr{c \to \alpha_1} + \kr{c \to \beta_1}}  \Phi^*_{\alpha} 
\nonumber\\
&=& \frac{ \kr{c \to \alpha_1}\Phi^*_{\alpha} }
{ \kr{c\to\alpha_1}\Phi^*_{\alpha} + \kr{c\to\beta_1}\Phi^*_{\beta} }.
\eea
This yield provides the residence time on the acceptor,
$r_{\alpha A} = \Gamma_{\alpha}^{-1} \Phi_{\alpha}$. Based
on the flux condition, we can now achieve the residence times of all other sites, recursively.

This technique can be generalized to treat roots with more than two branches by the modification of
the factor in the geometric series of (\ref{eq:fluxfromlinflux}) to include the probability
 of returning from additional branches.

Note that this approach allows us to  find the yield of
systems suffering from losses, which could physically correspond to particle trapping in the 
context of charge transfer of exciton recombination. 
Back to Fig. \ref{FigA}, we include a lossy process  of rate $\Gamma_s$ at site $c$, by
 modifying the `branch' at that site,
 $\kr{c \to \alpha_1} = \Gamma_s$, $\kl{\alpha_1\to c} = 0$.

\end{document}